\documentstyle[12pt,epsfig]{book}
\def \half{ \frac {1}{2} }
\def \be{\begin{equation}}
\def \ee{\end{equation}}
\def \bea{\begin{eqnarray}}
\def \eea{\end{eqnarray}}
\def \ba{\begin{array}}
\def \ea{\end{array}}
\def \non{\nonumber}
\def \h#1{{\hat{#1}}}
\def \ben{\begin{enumerate}}
\def \een{\end{enumerate}}

\def \ket#1{|#1\rangle}

\def \ra{\rightarrow}

\def \schr{Schr\"odinger }
\def \qg{quasigauge }
\def \Qg{Quasigauge }
\def \qe{quasiexactly }
\def \Qe{Quasiexactly }
\def \qes{QES }
\def \wkb{WKB }
\def \eps{\epsilon}
\def \sech{\mbox{sech}}
\def \csch{\mbox{csch}}

\begin{document}
%%%%%%%%%%%%%%%%%%%%%%%%%%%%%%%%%%%%%%%%%%%%%%%%%%%%%%%%%%%%
\pagestyle{empty}
\pagenumbering{roman}
%%%%%%%%%%%%%%%%%%%%%%%%%%%%%%%%%%%%%%%%%%%%%%%%%%%%%%%%%%%%
$~~~$
\newline
\newline
\newline
\newline
\newline
\newline
\newline
\newline
\newline
\newline
\centerline{\LARGE \bf QUASIEXACTLY SOLVABLE}
\newline
\newline
\centerline{\LARGE \bf POTENTIALS}
$~~~$
\newline
\newline
\newline
\newline
\newline
\newline
\newline
\newline
\centerline{\Large Nitsan Aizenshtark}
\newpage
$~~~$
%%%%%%%%%%%%%%%%%%%%%%%%%%%%%%%%%%%%%%%%%%%%%%%%%%%%%%%%%%%%
\newpage
$~~~$
\newline
\newline
\newline
\newline
\newline
\newline
\centerline{\LARGE \bf QUASIEXACTLY SOLVABLE}
\newline
\newline
\centerline{\LARGE \bf POTENTIALS}
\newline
\newline
\newline
\newline
\centerline{\large Research Thesis}
\newline
\newline
\newline
\newline
\newline
\newline
\newline
\newline
\centerline{Submitted in Partial Fulfillment of The Requirements for the Degree of}
\centerline{ Master of Science in Physics}
\newline
\newline
\newline
\newline
\centerline{\Large Nitsan Aizenshtark}
\newline
\newline
\newline
\newline
\newline
\newline
\newline
\centerline{Submitted to the Senate of the Technion - Israel Institute of Technology}
\newline
\newline
\newline
\newline
\centerline{TAMUZ 5763 $~~~~~$ HAIFA $~~~~~$  JULY 2003}
\newpage
$~~~$
%%%%%%%%%%%%%%%%%%%%%%%%%%%%%%%%%%%%%%%%%%%%%%%%%%%%%%%%%%%%
\newpage
$~~~$
\newline
\newline
The Research Thesis Was Done Under The Supervision of Prof. Moshe Moshe and Dr. Joshua Feinberg in the Faculty of Physics.
$~~~$
\newline
\newline
\newline
\newline
\newline
\newline
\newline
\newline
\newline
I would like to thank Prof. Moshe Moshe and Dr. Joshua Feinberg for their help and guidance during the preparation of this work.
\newline
\newline
\newline
I would like to thank Yael Roichman, Saar Rahav, Yair Srebro and Prof. Dov Levine for all their good advice.
\newline
\newline
\newline
I would like to thank my family and friends for their encouragement and support. 
\newline
\newline
\newline
I would like to thank Boaz for always being there for me.
\newline
\newline
\newline
\newline
\newline
\newline
\newline
\newline
\newline
\newline
\newline
THE GENEROUS FINANCIAL HELP OF THE TECHNION IS GRATEFULLY ACKNOWLEDGED.

%%%%%%%%%%%%%%%%%%%%%%%%%%%%%%%%%%%%%%%%%%%%%%%%%%%%%%%%%%%%
\tableofcontents
%%%%%%%%%%%%%%%%%%%%%%%%%%%%%%%%%%%%%%%%%%%%%%%%%%%%%%%%%%%%

%%%%%%%%%%%%%%%%%%%%%%%%%%%%%%%%%%%%%%%%%%%%%%%%%%%%%%%%%%%%
\listoftables
%%%%%%%%%%%%%%%%%%%%%%%%%%%%%%%%%%%%%%%%%%%%%%%%%%%%%%%%%%%%
\listoffigures
%%%%%%%%%%%%%%%%%%%%%%%%%%%%%%%%%%%%%%%%%%%%%%%%%%%%%%%%%%%%

\newpage
\pagestyle{plain}
\pagenumbering{arabic}
$~~~$
\newline
\newline
\newline
{\huge \bf Abstract}
\addcontentsline{toc}{chapter}{Abstract}
\newline
\newline
\newline
%%%%%%%%%%%%%%%%%%%%%%%%%%%%%%%%%%%%%%%%%%%%%%%%%%%%%%%%%%%%
Quantum mechanical potentials can be divided according to their solvability into three categories.
The best known category is exactly solvable potentials (e.g. the harmonic oscillator and the hydrogen atom), which can be solved analytically.
This means that all the energy levels and all the wave functions can be calculated in a finite number of algebraic steps.
The second category is non-solvable potentials (e.g. the quartic potential), for which it is only possible to approximate the solution using various analytical and numerical methods.
The third and least known category is \qe solvable (QES) potentials, for which only part of the energy spectrum and the corresponding wave functions can be analytically calculated.

The feature distinguishing \qes potentials from non solvable potentials is that their Hamiltonian may be represented as a block-diagonal matrix with at least one finite block.
Such a finite block may always be diagonalized, thus the energy levels and wave functions corresponding to it can always be calculated.
Block diagonal Hamiltonians can be created using bilinear combinations of Lie group generators.
The exactly solvable potentials may be treated as a sub-group of the \qes potentials since it is always possible to diagonalize their Hamiltonian.

This work deals with the solution of one dimensional \qes potentials. 
It also includes one dimensional supersymmetric \qes potentials which are potentials that can be represented by a $ 2 \times 2 $ matrix and their Hamiltonian operates on a wave function which is a two component spinor of one spatial variable.

First, expanding the results of existing works, some properties of the exact solutions of \qes potentials are shown to hold for the general case.
One interesting property of some \qes potentials (which is used in the second part of this work) is the simple connection between the wave functions and the initial value solution of the \schr equation.
The initial value solution is the generating function of the secular polynomials, the zeros of which are the \qe solvable part of the energy spectrum for the potential.
This property is shown to hold for a class of \qes and exactly solvable potentials and not only for specific examples.

Next, the secular polynomials and the \qes part of the energy spectrum are approximated using a new technique which is based on the \wkb approximation.
The approximation is useful because calculating the energy levels exactly becomes more and more complicated as the size of the finite block in the Hamiltonian grows.
The initial value solution of the \schr equation is calculated using the \wkb approximation.
Then approximate secular polynomials and energy levels are derived from the \wkb solutions.
In order to simplify this process saddle point approximation is used to obtain the secular polynomials and the energy levels from the \wkb approximation.
This calculation has to be carried out separately for every potential.
The process is demonstrated for the harmonic oscillator and the sextic potential
and the quality of the results is discussed.
The combined \wkb and saddle point approximations give a simple evaluation of the energy levels of \qes potentials with a generating function that solves the \schr equation.
%%%%%%%%%%%%%%%%%%%%%%%%%%%%%%%%%%%%%%%%%%%%%%%%%%%%%%%%%%%%
\newpage
$~~~$
\newline
\newline
\newline
{\huge \bf List of Symbols}
\addcontentsline{toc}{chapter}{List of Symbols}
\newline
\newline
\newline
%%%%%%%%%%%%%%%%%%%%%%%%%%%%%%%%%%%%%%%%%%%%%%%%%%%%%%%%%%%%
\begin{tabular}{ll}
\hline \\
symbol & meaning  \\
 \\
\hline \\
$ a(x) $ & imaginary phase and exponential behavior of the wave function\\
$ {\cal A}(x) $ & derivative of $ a(x) $ and \qg field\\
$ C_k $ & wave function coefficient\\
$ c_{ab} , \; c_a $ & coefficients of the generators in $ H_G $\\
$ D_k $ & even wave function coefficient\\
$ d_k $ & odd wave function coefficient\\
$ d_k $ & lower wave function coefficient\\
$ E $ & energy\\
$ E_n $ & energy eigenvalue\\
$ E_\ast $ & highest energy of the \qes sector\\
$ H $ & Hamiltonian\\
$ H_G $ & \qg Hamiltonian\\
$ j $ & the group parameter\\
$ J $ & $ u(1) $ group generator\\
$ {\bf M}_{2j+1} $ & coefficient matrix\\
$ M_{(n,k)} $ & minor of the term in the $ n+1 $ row and the $ k+1 $ column of $ {\bf M} $\\
$ {\bf m}_{2j} $ & minor of coefficient matrix\\
$ {\bf N}_{k} $ & minor of coefficient matrix\\
$ P_k(E) $ & secular polynomials\\
$ p_i(\xi) $ & the coefficients of the $ \xi $ derivatives in $ H_G $\\
$ Q_k(E) $ & even secular polynomials\\
$ q_k(E) $ & odd secular polynomials\\
$ Q_\alpha , \;  \bar{Q}_\alpha $ & group generators\\
$ R_j $ & the $ j $th representation of a group\\
$ S_\alpha $ & phase of Cauchy integral\\
$ s $ & scaled $ x $\\
\end{tabular}
\newpage
$~~~~$
\newline
\newline
\begin{tabular}{ll}
\hline \\
symbol & meaning  \\
 \\
\hline \\
$ s_\beta $ & saddle point of  $ I_\alpha $\\
$ T^a $ & group generator\\
$ t $ & scaled $ y $\\
$ u_k $ & lower wave function coefficient\\
$ V(x) $ & potential\\
$ v(s) $ &  scaled $ V(x) $\\
$ \alpha_k $ & a coefficient in the recursion of the wave function coefficients\\
$ \beta_k $ & a coefficient in the recursion of the wave function coefficients\\
$ \gamma_k $ & a coefficient in the recursion of the wave function coefficients\\
$ \delta_k $ & a coefficient in the recursion of the wave function coefficients\\
$ \Delta_k $ & the proportion relation of $ C_k $ to $ P_k(E) $\\
$\Delta V(x) $ & free function in $ H_G $\\
$ \eps_k $ & a coefficient in the recursion of the wave function coefficients\\
$ \eps $ & scaled $ E $\\
$ \kappa $ & Cramer's determinant\\
$ \xi $ & the group variable\\
$ \psi $ & the wave function\\
$ \tilde{\psi} $ & the pre-exponential function\\
$ \Psi $ & the initial value solution of the \schr equation\\
\end{tabular}
\newpage

%%%%%%%%%%%%%%%%%%%%%%%%%%%%%%%%%%%%%%%%%%%%%%%%%%%%%%%%%%%%
\chapter{Introduction}
%%%%%%%%%%%%%%%%%%%%%%%%%%%%%%%%%%%%%%%%%%%%%%%%%%%%%%%%%%%%

\label{intro}
In this chapter the basic properties of the \qes potentials are described and the simplest example (the sextic potential) is analyzed.
\Qe solvable (QES) potentials are distinguished by the fact that a finite number of wave functions and energy levels can be analytically calculated \cite{shifmanreview}.
Hamiltonians corresponding to \qes potentials can be constructed from bilinear combinations of Lie group generators.
Hamiltonians are second order differential operators, and does not have a first order derivative term, whereas a bilinear combination of Lie group generators does.
By transforming the Hamiltonian into a \qg invariant Hamiltonian, the two representations (Hamiltonian and Lie group generators) of the \schr equation can be made equivalent through a change of variables.
By choosing the basis of the wave functions as the direct sum of the representation of the Lie group and an arbitrary set from the orthogonal space, the Hamiltonian becomes block-diagonal with at least one finite block.
The solutions corresponding to this block may be calculated exactly.
This work mainly deals with one dimensional \qes potentials which are generated by the $ sl(2) $ Lie group generators.
This work is within the Lie-algebraic formalism \cite{turbiner88}.
An analytic formalism is summarized in the review article \cite{ushveridze89}.

%%%%%%%%%%%%%%%%%%%%%%%%%%%%%%%%%%%%%%%%%%%%%%%%%%%%%%%%%%%%
\section{The \Qg Symmetry}
%%%%%%%%%%%%%%%%%%%%%%%%%%%%%%%%%%%%%%%%%%%%%%%%%%%%%%%%%%%%

The \schr equation (defined for example in \cite{gasiorowicz}) can be solved in numerous ways, exact and approximate.
This equation is gauge invariant even though the Hamiltonian is not.
A gauge invariant Hamiltonian can be created by replacing the derivatives by covariant derivatives.
\Qg invariance is defined as invariance under imaginary phase gauge transformations.
The \qg Hamiltonian is defined as a \qg invariant operator which is derived from the Hamiltonian by a real correction to the derivative (an imaginary gauge field).
The \schr equation is not \qg invariant but if the Hamiltonian is replaced by the \qg Hamiltonian we can achieve a \qg invariant formalism.

For all potentials, the solution of the \schr equation 
\be
\label{schr}
H \psi (x) = \left[ - \half \frac{\partial^2}{\partial x^2} + V(x) \right] \psi (x) = E \psi (x)
\ee
can be written as the product
\be
\label{exponential}
\psi(x) = e^{-a(x)} \tilde{\psi}(x)
\ee
where $ a(x) $ is the imaginary phase, $ e^{-a(x)} $ is the exponential factor and $ \tilde{\psi}(x) $ is the pre-exponential function.
Substituting $ \psi(x) $ in the \schr equation leads to a new equation for the pre-exponential function $ \tilde{\psi}(x) $
\be
\label{quasigaugeschr}
H_G \tilde{\psi} =  \left[ - \half \left( \frac{\partial}{\partial x_i} - {\cal A}_i(x) \right) ^2 + V(x) \right] \tilde{\psi}  = E \tilde{\psi}
\ee
where $ {\cal A}_i(x) = \frac{d}{dx_i}a(x) $.
The operator $ H_G $ is called the \qg Hamiltonian and the new equation is invariant under the \qg transformation
\be
\tilde{\psi} \ra \tilde{\psi} e^{-b(x)} , \; {\cal A}_i \ra {\cal A}_i - \frac{\partial b}{\partial x_i}.
\ee
The eigenvalues (energy levels) of equation (\ref{quasigaugeschr}) and of the \schr equation (\ref{schr}) are identical.
The eigenfunctions of the two equations are related through (\ref{exponential}).

%%%%%%%%%%%%%%%%%%%%%%%%%%%%%%%%%%%%%%%%%%%%%%%%%%%%%%%%%%%%
\section{The \Qg Hamiltonian}
%%%%%%%%%%%%%%%%%%%%%%%%%%%%%%%%%%%%%%%%%%%%%%%%%%%%%%%%%%%%

\qes potentials are distinguished by the fact that their \qg Hamiltonian may be written as a block diagonal matrix with at least one finite block.
\be
H_G = \left( \ba{c|c}
\ba{cccc}
h_{1,1} & h_{1,2} & h_{1,3} & \ddots \\
h_{2,1} & h_{2,2} & \ddots & h_{N-2,N} \\
h_{3,1} & \ddots & h_{N-1,N-1} & h_{N-1,N} \\
\ddots & h_{N,N-2} & h_{N,N-1} & h_{N,N} \\
\ea & 0 \\ \hline
0 & h_{ij} \ne 0
 \ea \right).
\ee
The energy levels and wave functions corresponding to the finite block can always be calculated.
\Qg Hamiltonians can be generated by taking bilinear combinations of Lie group generators
\be
H_G = \sum_{a,b} c_{ab}T^a T^b + \sum_{a} c_a T^a
\ee
in a basis $ \{ \ket{\tilde{\psi}} \} $ which is the direct sum of the $ j $th representation $ R_j $ of the Lie group and an arbitrary set from the orthogonal space
\be
\{ \ket{\tilde{\psi}} \} = \{ \mbox{elements of } R_j + \mbox{arbitrary set from orthogonal space} \}.
\ee
A first order differential representation is chosen for the generators so the \qg Hamiltonian is a second order differential operator equivalent to (\ref{quasigaugeschr}) up to a change of variables.

Many examples of \qes potentials are found in the literature: one dimensional potentials \cite{turbiner88}, one dimensional radial potentials \cite{shifman89b}, \qes potentials in higher dimensions, spatial and Grassmann (supersymmetric) \cite{shifmanturbiner89,shifman89a,gonzalez94,shifman90,brihaye,finkel97}.
Many particle systems of \qes potentials are analyzed for example in \cite{shifman99b,mayer00}.
Because the \qes potentials are an intermediate class between the exactly solvable potentials and the non solvable potentials they were used as a tool in many calculations, for example \cite{cicuta96,tymczak97}.

%%%%%%%%%%%%%%%%%%%%%%%%%%%%%%%%%%%%%%%%%%%%%%%%%%%%%%%%%%%%
\section{One-Dimensional \qes Potentials}
%%%%%%%%%%%%%%%%%%%%%%%%%%%%%%%%%%%%%%%%%%%%%%%%%%%%%%%%%%%%

This work concentrates on one-dimensional \qes potentials.
This is because in one-dimensional quantum mechanics the ordering of the energy levels and wave functions is known (for a review of oscillation theorem see appendix \ref{oscillation}).
The one-dimensional \qes potentials discussed in this work are related to the $ sl(2) $ Lie algebra \cite{shifmanreview} defined by the commutation relations 
\be
[T^+,T^-]=2T^0 \; , \; [T^+,T^0]=-T^+ \; , \; [T^-,T^0]=+T^-.
\ee
This algebra may by represented by $ 2 \times 2 $ matrices that are non-singular under matrix multiplication with determinant $ +1 $ \cite{grouptheory,liegroups}.
The $ sl(2) $ group representations are characterized by a single parameter $ j $, where the $ j $th representation is of dimension $ 2j+1 $.
The differential representation of the generators corresponding to $ j $ is
\bea
T^+ & = & 2j\xi-\xi^2 \frac{d}{d\xi} \label{tplus} \\
T^0 & = & -j+\xi\frac{d}{d\xi} \label{tzero} \\
T^- & = & \frac{d}{d\xi} \label{tminus}.
\eea
The $ j $th polynomial representation is composed of powers of the group variable, from zero to $ 2j $
\be
\{ R_j \} = \{ \xi ^0, \xi ^1,...,\xi ^{2j}\}.
\ee
The basis of the solution space is $ \{ \xi ^0, \xi ^1,...,\xi ^{2j},\tilde{\psi}_{2j+2},\tilde{\psi}_{2j+3},...\} $ where $ \{ \tilde{\psi}_{2j+k} \} $ is an arbitrary set that spans the orthogonal space.
The pre-exponential functions $ {\tilde \psi}(x) $ of the \qes sector are polynomials of degree $ 2j $ and the general wave function, which solves the \schr equation (\ref{schr}), has the form 
\be
\label{generalwavef}
\psi(x)= e^{-a(x)} \tilde{\psi} = e^{-a(x)} \sum _{k=0}^{2j} C_k \xi^k(x). 
\ee
Writing $H_G$ as an explicit expression in the variable $\xi$ results in
\bea
H_G & = & \sum_{a,b= \pm ,0} c_{ab}T^a T^b + \sum_{a= \pm ,0} c_a T^a \non \\
& = & - \half p_4(\xi)\frac{d^2}{d \xi^2}+p_3(\xi)\frac{d}{d\xi}+p_2(\xi)
\label{qghamiltonian}
\eea
where $ p_i(\xi) $ is a polynomial of degree $ i $.
Without loss of generality the coefficients $ c_{ab} $ can be taken to be symmetric $ (c_{ab}=c_{ba}) $.
The polynomials $ p_4(\xi) , \; p_3(\xi) , \; p_2(\xi) $ under this assumption are given by
\bea
\label{hgp}
p_4(\xi) & = & -2 [ c_{++} \xi ^4 - 2c_{+0} \xi ^3 + ( c_{00} - 2c_{+-} ) \xi ^2 + 2c_{0-} \xi + c_{--} ] \non \\
p_3(\xi) & = & -2(2j-1)c_{++}\xi^3 + (3(2j-1)c_{+0}-c_+)\xi^2 \non \\
& + & (-(2j-1) (c_{00}-2c_{+-})+c_0)\xi + (-(2j-1)c_{0-}+c_-)  \non \\
p_2(\xi) & = & 2j(2j-1)c_{++}\xi^2 + (-2j(2j-1)c_{+0}+2jc_+)\xi.
\eea 
The transformation between the \schr picture and the \qg picture is found by comparing the \qg Hamiltonian in $ \xi $ (\ref{qghamiltonian}) to the \qg Hamiltonian in $ x $ (\ref{quasigaugeschr}).
\be
\label{qghamx}
H_G = - \half \left( \frac{\partial}{\partial x} - {\cal A}(x) \right) ^2 + V(x).
\ee
The change of variables $ x(\xi) $ is found by comparing the coefficients of the second derivatives
\be
\label{xi}
x(\xi) = \int^\xi \frac {d\xi'}{\sqrt{p_4(\xi')}}.
\ee
Comparing the coefficients of the first derivatives yields the \qg field $ {\cal A} (x) $
\be
\label{aderivative}
{\cal A} (x) = \left. \frac{p'_4/4 +p_3}{\sqrt{p_4}} \right|_{\xi=\xi(x)}
\ee
and the exponential behavior $ a(x) $
\be
\label{aphase}
a(x) = \int {\cal A}(x)dx = \int^{\xi(x)} \frac {p'_4/4 +p_3}{p_4}d\xi.
\ee
$ \Delta V(x) $ is defined as
\be
\label{deltav}
\Delta V(x) = \left. p_2(\xi) \right|_{\xi=\xi(x)}
\ee
The free functions are compared to give the \schr potential $ V(x) $
\be
\label{potential}
V(x) = \Delta V(x) + \half {\cal A}^2(x) - \half {\cal A}'(x).
\ee
It is now possible to take any bilinear combination of $ sl(2) $ generators and find the corresponding \qes potential \cite{shifmanreview,turbiner88}.

The \schr equation of any \qes potential can now be replaced using (\ref{generalwavef}) to get a set of $ 2j+1 $ equations for the wave function coefficients $ \{ C_k \}_{k=0}^{2j} $.
This set of homogeneous equations has a non-trivial solution if the determinant of the coefficient matrix (the secular polynomial) is zero.
This is the quantization condition of the energy levels.
The generators can be defined differently \cite{debergh00} so that the Hamiltonian is a linear combination of generators but that does not change the range of \qes potential obtained by this formalism and their solutions.
The properties of the Bender-Dunne polynomials and other sets of polynomials that are the wave function coefficients of the \qes potentials are analyzed in many articles.
The orthogonality of the wave function coefficients is discussed for example in \cite{finkel96,krajewska96b}.

%%%%%%%%%%%%%%%%%%%%%%%%%%%%%%%%%%%%%%%%%%%%%%%%%%%%%%%%%%%%
\subsection{Example 1 - The Sextic Potential}
%%%%%%%%%%%%%%%%%%%%%%%%%%%%%%%%%%%%%%%%%%%%%%%%%%%%%%%%%%%%

\label{simplest}
The simplest and best known example of a \qes potential is the sextic potential with a quantized coefficient before the harmonic term.
\be
\label{simplestpotential}
V(x)=\frac{1}{2}x^6-\frac{1}{2}(8j+3)x^2.
\ee
\begin{figure}[h]
\epsfig{file=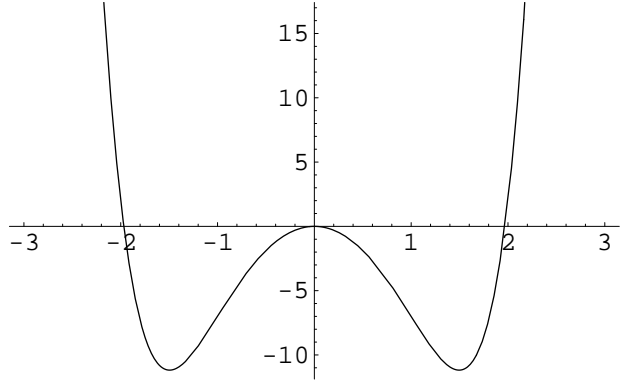}
\caption{\label{simplestp} The sextic potential with $ j = \frac{3}{2} $.}
\end{figure}
This example has been analyzed in many articles.

This potential has an $ x \ra -x $ symmetry so a natural choice for a new variable is $ \xi(x) = x^2 $.
In $ \xi $ the kinetic term of the Hamiltonian is
\be
- \half \frac{d^2}{dx^2} = - 2 \frac{d^2}{d\xi^2} - \frac{d}{d\xi}
\ee
which can be written in the generators of $ sl(2) $ (\ref{tplus}-\ref{tminus}) as
\be
- \half \frac{d^2}{dx^2} = -2T^0T^- - (2j+1)T^-.
\ee
The potential energy term can be any linear combination of the generators.
The sextic potential (\ref{simplestpotential}) is the result of choosing the \qg Hamiltonian
\be
\label{simplesthg}
H_G = -2T^0T^- - (2j+1)T^- - 2T^+ = -2\xi \frac{d^2}{d \xi ^2} + (2 \xi ^2 -1)\frac{d}{d \xi}-4j \xi.
\ee
Using the definitions of equations (\ref{xi}-\ref{aphase}) the wave functions have the form
\be
\psi(x)= e^{-x^4/4}\sum_{k=0}^{2j}C_kx^{2k}
\ee
In \cite{benderdunnemoshe} the following recursion relation for the coefficients of the wave functions
\be
\label{simplestr}
0=2(2j+1-k)C_{k-1} + EC_k + (k+1)(2k+1)C_{k+1}
\ee
with the boundary conditions
\be
C_{-1}=0 , \; C_{2j+1}=0
\ee
is derived.
\begin{figure}[h]
\epsfig{file=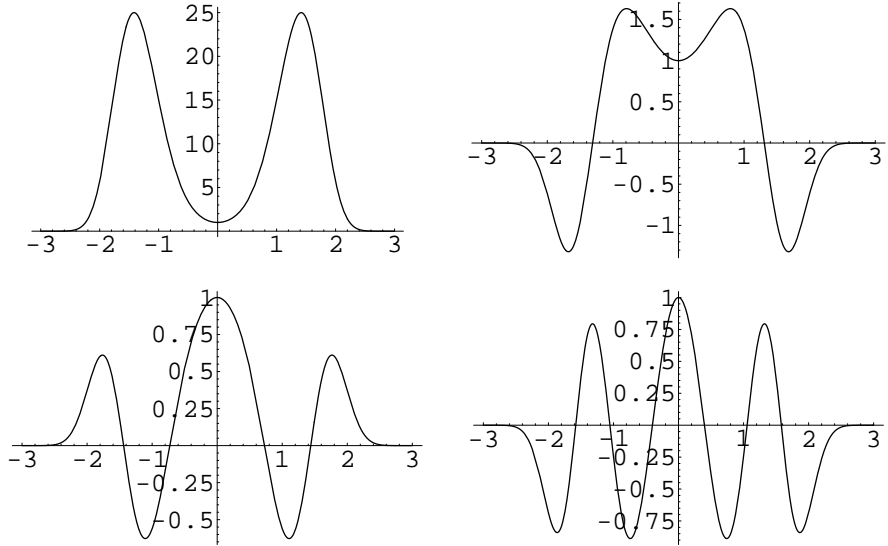}
\caption{The wave functions of the sextic potential with $ j = \frac{3}{2} $.}
\end{figure}
In \cite{benderdunne96} solutions of the form
\be
\label{psit}
\psi(x) = e^{-x^4/4} \sum_{k=0}^{\infty} \frac{(-2)^k}{(2k)!}P_k(E) x^{2k}
\ee
are substituted in the \schr equation to give the recursion relation
\be
P_{k+1}(E)=EP_k(E)-2k(2k-1)(2j+1-k)P_{k-1}(E).
\ee
For $ k=2j+1 $ the last coefficient of the recursion vanishes and
\be
\label{factor}
P_{2j+1+n}(E)=P_{2j+1}(E)Q_n(E)
\ee
which means that all the polynomials $ P_{2j+1+n}(E) $ have the same zeros as $ P_{2j+1}(E) $.
This leads to two conclusions. 
First, for all the zeros of $ P_{2j+1}(E) $ the wave function (\ref{psit}) is normalizable because the sum is truncated.
Second, the polynomial $ P_{2j+1}(E) $ is the secular polynomial of the set of equations (\ref{simplestr}) if the initial conditions are chosen as
\be
P_0=1 , \; P_{1}=E,
\ee
And the initial value solution of the \schr equation (\ref{psit}) is the generating function of the secular polynomials.
\begin{table}[h]
\begin{tabular}{|l||l|l|} \hline
$j$ & $P_{2j+1}(E)$ & $E$ \\ \hline \hline
$0$ & $E$ & $0$ \\ \hline
$\half$ & $E^2 - 2$ & $\pm \sqrt{2}$ \\ \hline
$1$ & $E^3-16E$ & $0, \pm 4$ \\ \hline
$\frac{3}{2}$ & $E^4-60E^2+180$ & $\pm 1.77966, \pm 7.53875$ \\ \hline
$2$ & $E^5-160E^3+2944E$ & $0, \pm 4.60568, \pm 11.7808$ \\ \hline
$\frac{5}{2}$ & $E^6-350E^4+20716E^2-81000$ & $\pm 2.05066, \pm 8.35389, \pm 16.6135$ \\ \hline
$3$ & $E^7-672E^5+95616E^3-2045952E$ & $0, \pm 5.09375, \pm 12.7813, \pm 21.9703$ \\ \hline
\end{tabular}
\caption{\label{x^6} The secular polynomials and \qes sector energy levels of the sextic potential.}
\end{table}
The first polynomials and energies are listed in table (\ref{x^6}).

The sextic potential has an additional symmetry, known as the energy reflection symmetry or self duality.
This symmetry was first analyzed and used in \cite{benderdunnemoshe} to explain the large $ j $ approximation of the highest energy level of the \qes sector $ E_*=\frac{16}{3} \sqrt{\frac{2}{3}} $ as the reflection of the lowest energy at the bottom of the potential well.
More potentials were found to be self dual. 
Other potentials were found to satisfy a weaker form of the symmetry which is known as duality \cite{krajewska96a,shifman99a,dunne02}.

%%%%%%%%%%%%%%%%%%%%%%%%%%%%%%%%%%%%%%%%%%%%%%%%%%%%%%%%%%%%
\subsubsection{Parity of the Wave Functions}
%%%%%%%%%%%%%%%%%%%%%%%%%%%%%%%%%%%%%%%%%%%%%%%%%%%%%%%%%%%%

The solutions (\ref{psit}) are even functions.
The odd wave functions of the potential (\ref{simplestp}) are not a part of the \qe solvable sector.
A very similar potential with an odd \qes sector can be derived by transforming the \qg Hamiltonian (\ref{simplesthg}) by:
\be
H_G \rightarrow H_G -2T^-.
\ee
This new \qg Hamiltonian is related through (\ref{potential}) to the potential 
\be
V(x)=\frac{1}{2}x^6-\frac{1}{2}(8j+5)x^2,
\ee
which is similar to the original potential (\ref{simplestp}) but has odd \qes wave functions. 
The wave functions of this \qg Hamiltonian have the form
\be
\psi(x)=e^{-x^4/4}x\sum_{k=0}^{2j}C_kx^{2k}
\ee
and the coefficients $ C_k $ are connected by the recursion relation 
\be
0=2(2j+1-k)C_{k-1} + EC_k + (k+1)(2k+1)C_{k+1}
\ee
which is different from (\ref{simplestr}) but has the same initial conditions.
The even wave functions of this potential are not a part of the \qe solvable sector.

%%%%%%%%%%%%%%%%%%%%%%%%%%%%%%%%%%%%%%%%%%%%%%%%%%%%%%%%%%%%
\section{Outline}
%%%%%%%%%%%%%%%%%%%%%%%%%%%%%%%%%%%%%%%%%%%%%%%%%%%%%%%%%%%%

This work concentrates on one dimensional \qes potentials.

In chapter (\ref{exact}) the \qes potentials and the various properties of their exact solutions are discussed.
The wave function coefficients are given a new explicit form in terms of the coefficient matrix.
A new formula for calculating the secular polynomials is derived.
Several special cases are discussed including supersymmetric potentials.

In Chapter (\ref{approximations}) of this work, a new technique based on the \wkb approximation is developed, and is used to approximate the highest energy levels for large $ j $.
This technique combines the initial value \wkb approximation with the saddle point approximation.
It applies to all cases that have a generating function of the secular polynomials that is the initial value solution of the \schr equation.
Two explicit examples of using this technique are given.

In chapter (\ref{summary}) of this work the results and conclusions are summarized and discussed.

%%%%%%%%%%%%%%%%%%%%%%%%%%%%%%%%%%%%%%%%%%%%%%%%%%%%%%%%%%%%
\chapter{Exact Solutions}
%%%%%%%%%%%%%%%%%%%%%%%%%%%%%%%%%%%%%%%%%%%%%%%%%%%%%%%%%%%%

\label{exact}
\qes potentials are distinguished by having a finite number of exact solutions. The number of the exact solutions is potential dependent. This attribute is related to the symmetry of the Hamiltonian.
In order to explicitly see the underlying symmetry it is more convenient to work with an alternative operator: the \qg Hamiltonian, $H_G$, which is symmetric under \qg transformations.
General exact solutions are found and the methods are applied to several one-dimensional and one-dimensional SUSY potentials.

%%%%%%%%%%%%%%%%%%%%%%%%%%%%%%%%%%%%%%%%%%%%%%%%%%%%%%%%%%%%
\section{General Solutions}
%%%%%%%%%%%%%%%%%%%%%%%%%%%%%%%%%%%%%%%%%%%%%%%%%%%%%%%%%%%%

\label{general}
For the one-dimensional \qes potentials it is possible to replace the \schr equation with the equation
\be
\label{qesschr}
H_G{\tilde \psi} = \left[ \sum_{a,b= \pm ,0} c_{ab}T^a T^b + \sum_{a= \pm ,0} c_a T^a \right]{\tilde \psi} = E {\tilde \psi}
\ee
where $ T^a , \; a=\pm,0 $ are the generators of the $ sl(2) $ Lie group.
In the $ j $th polynomial representation the solutions $ \tilde \psi $ are given by a power series in the group variable $ \xi $
\be
\label{wave}
{\tilde \psi} = \sum _k C_k \xi^k 
\ee
Equation (\ref{qesschr}) can now be transformed into a recursion relation for the coefficients $ C_k $.
For the most general one-dimensional case this relation is a 5 term recursion relation:
\bea.
\label{generalrecursion}
0 & = & -c_{++}(2j+2-k)(2j+1-k) C_{k-2} \nonumber \\
& - & \left[ c_{+0}(2k-2j-1)+c_+ \right] (2j+1-k)C_{k-1} \nonumber \\
& + & \left[ E +(c_{00}-2c_{+-})(2j-k)k-c_0k \right]C_{k} \nonumber \\
& - & \left[ c_{0-} (2k-2j+1)+c_- \right] (k+1) C_{k+1} \nonumber \\
& - & c_{--}(k+1)(k+2) C_{k+2}.
\eea
If we define $ \alpha_k , \; \beta_k , \; \gamma_k , \; \delta_k, \; \epsilon_k $ to be the functions
\bea
\label{abcde}
\alpha_k & = & -c_{++}(2j+2-k)(2j+1-k) \nonumber \\
\beta_k & = & - \left[ c_{+0}(2k-2j-1)+c_+ \right] (2j+1-k) \nonumber \\
\gamma_k & = & E +[(c_{00}-2c_{+-})(2j-k)-c_0]k \nonumber \\
\delta_k & = & - \left[ c_{0-}(2k-2j+1) + c_- \right] (k+1) \nonumber \\
\epsilon_k & = & -c_{--}(k+1)(k+2)
\eea
then the recursion relation can be written as
\be
0=\alpha_k C_{k-2} + \beta_k C_{k-1} + \gamma_k C_{k} + \delta_k C_{k+1} + \eps_k C_{k+2}.
\ee
Wave functions are boundary value solutions of the \schr equation.
This remains unchanged though it is now a difference equation and not a differential equation.
In order to truncate the sum (\ref{wave}) and ensure the normalizability of the \schr wave function the boundary conditions are chosen to be
\be
C_{-2}=0 , \; C_{-1}=0 , \; C_{2j+1}=0 , \; C_{2j+2}=0.
\ee
This, with the special form of the coefficients (\ref{abcde}) results in the normalizable wave function
\be
\psi(x)= e^{-a(x)} \sum _{k=0}^{2j} C_k \xi^k(x),
\ee
which has no more than $ 2j+1 $ zeros.

The set of equations giving the coefficients $ \{ C_k \}_{k=0}^{2j} $ can be written in matrix form as
\be
{\bf M}_{2j+1} \vec{C} = 0
\ee
where $ {\bf M}_{2j+1} $ is the ${(2j+1) \times (2j+1)}$ coefficient matrix:
\be
\label{matrix}
{\bf M}_{2j+1} = \left( \ba{cccccccc}
\gamma_0 & \delta_0 & \epsilon_0 & 0 & 0 & 0 & \ddots \\
\beta_1 & \gamma_1 & \delta_1 & \epsilon_1 & 0 & \ddots & 0 \\
\alpha_2 & \beta_2 & \gamma_2 & \delta_2 & \ddots & 0 & 0 \\
0 & \alpha_3 & \beta_3 & \ddots & \delta_{2j-3} & \epsilon_{2j-3} & 0 \\
0 & 0 & \ddots & \beta_{2j-2} & \gamma_{2j-2} & \delta_{2j-2} & \eps_{2j-2} \\
0 & \ddots & 0 & \alpha_{2j-1} & \beta_{2j-1} & \gamma_{2j-1} & \delta_{2j-1} \\
\ddots & 0 & 0 & 0 & \alpha_{2j} & \beta_{2j} & \gamma_{2j}
\ea
\right)
\ee
($ \alpha_k , \; \beta_k , \; \gamma_k , \; \delta_k, \; \epsilon_k $ as defined in (\ref{abcde})), and $ \vec C $ is the vector of variables $ \vec{C} $: 
\be
\vec{C} =  \left( \ba{c}
C_0 \\ C_1 \\ C_2 \\ \vdots \\ C_{2j-2} \\ C_{2j-1} \\ C_{2j}
\ea \right).
\ee
The orthogonality of the wave function coefficients as functions of the energy $ E $ is analyzed in \cite{finkel96,krajewska96b} and will not be addressed in this work.
This set of homogeneous equations has a non trivial solution if and only if the determinant of the coefficient matrix vanishes
\be
\label{quantization}
P_{2j+1}(E) = \det {\bf M}_{(2j+1)} = 0.
\ee
This is the quantization condition for the energies of the system.
The polynomial $ P_{2j+1}(E) $ is known as the secular polynomial (or secular determinant).

%%%%%%%%%%%%%%%%%%%%%%%%%%%%%%%%%%%%%%%%%%%%%%%%%%%%%%%%%%%%
\subsection{The Wave Function Coefficients}
%%%%%%%%%%%%%%%%%%%%%%%%%%%%%%%%%%%%%%%%%%%%%%%%%%%%%%%%%%%%

The wave function coefficients can be given a general form in terms of the minors of $ {\bf M}_{2j+1} $.
This calculation may be generalized to give a solution for a set of $ n $ homogeneous equations in $ n $ variables.
The matrix representation of the set of equations has the general form
\be
\label{hset}
{\bf M}_{2j+1} \; \vec{C}_{2j+1} =
\left( \ba{ccccc}
\gamma_0 & \delta_0 & \epsilon_0 & 0 & \ddots \\
\beta_1 & \gamma_1 & \delta_1 & \ddots & 0 \\
\alpha_2 & \beta_2 & \ddots & \delta_{2j-2} & \epsilon_{2j-2} \\
0 & \ddots & \beta_{2j-1} & \gamma_{2j-1} & \delta_{2j-1} \\
\ddots & 0 & \alpha_{2j} & \beta_{2j} & \gamma_{2j}
\ea
\right) \left( \ba{c}
C_0 \\ C_1 \\ \vdots \\ C_{2j-2} \\ C_{2j-1} \\ C_{2j}
\ea \right) = 0.
\ee
This set has a non-trivial solution if the determinant of the coefficient matrix vanishes (\ref{quantization}).
In such cases the system of equations is linearly dependent and there are less linearly independent equations than variables.
Since this is a one dimensional quantum system there is no degeneracy in the energy levels, so we can expect that there is only one free parameter and one extra equation.
For a specific \qes potential the coefficient $ C_{2j} $ can never vanish because it will give a solution with less zeros than is allowed by oscillation theory (see appendix \ref{oscillation}). This makes $ C_{2j} $ a natural choice for the free parameter.
The last equation is chosen to be redundant.
The system of the first $ 2j $ equations ($ 0 \le k \le (2j-1) $) of the system (\ref{hset}) for the variables $ C_0,...,C_{2j} $ is an inhomogeneous system given by
\be
 \left( \ba{cccc}
\gamma_0 & \delta_0 & \epsilon_0 & 0 \\
\beta_1 & \gamma_1 & \delta_1 & \ddots \\
\alpha_2 & \beta_2 & \ddots & \delta_{2j-2} \\
0 & \ddots & \beta_{2j-1} & \gamma_{2j-1}
\ea
\right)
 \left( \ba{c}
C_0 \\ C_1 \\ \vdots \\ C_{2j-1}
\ea \right) =  \left( \ba{c}
\vdots \\ 0 \\ \epsilon_{2j-2}C_{2j} \\ \delta_{2j-1}C_{2j}
\ea \right),
\ee
which can be written in matrix notation as
\be
{\bf m}_{2j} \; \vec{C}_{2j}= \vec{b}_{2j}.
\ee
The solution of this inhomogeneous system is given by Cramer's rule:
\be
\label{ihsol}
C_k = \frac{\det \left( \kappa_k \right) }{\det {\bf m}_{2j}},
\ee
where $ \kappa_k $ is the determinant of a matrix that is obtained from $ {\bf m}_{2j} $ by replacing the $ k^{th} $ column with the vector $ \vec{b}_{2j} $:
\be
(\kappa_k)_{n,m} = \left\{ \ba {ll}
({\bf m}_{2j})_{n,m} & m \ne k \\
(\vec{b}_{2j})_n & m = k.
\ea \right.
\ee
We have chosen $ C_{2j} = 1 $ in (\ref{ihsol}), changing this choice amounts to a multiplication of the wave function by a constant and does not change the following conclusions since none of these solutions are normalized.
We want to express the solution (\ref{ihsol}) in terms of the minors of the original matrix  (\ref{matrix}).
It is easy to see that $ \kappa_k $ is identical to $ (-1)^{2j-k} \det M_{(2j,k)}$ where $ M_{(n,k)} $ is the minor obtained by eliminating the $ n+1 $ row and the $ k+1 $ column of (\ref{matrix}). The solution for $ C_0,...,C_{2j} $, in terms of the minors of $ {\bf M}_{2j+1} $, is
\be
\label{hsol}
C_k=\frac{(-1)^{2j-k} \det M_{(2j,k)} }{\det M_{(2j,2j)}}.
\ee
The free parameter $ C_{2j} $ can also be included by this determinant ratio since
\be
\label{hsollast}
C_{2j} = 1 = \frac{\det M_{(2j,2j)}}{\det M_{(2j,2j)}}.
\ee
It can be proved that the boundary condition $ C_{2j+1}=0 $ is satisfied automatically since the form of the recursion relation (\ref{generalrecursion}) for $ k=2j-1 $
\bea
0 & = & \alpha_{2j-1}C_{2j-3}+\beta_{2j-1}C_{2j-2}+(E+\gamma_{2j-1})C_{2j-1} \non \\ 
& + & \delta_{2j-1}C_{2j}+\epsilon_{2j-1}C_{2j+1}
\eea
leads to
\bea
C_{2j+1} & = & \frac{-1}{\epsilon_{2j-1}}[\alpha_{2j-1}C_{2j-3}+\beta_{2j-1}C_{2j-2} \non \\
& + & (E+\gamma_{2j-1})C_{2j-1}+\delta_{2j-1}C_{2j}].
\eea
Using (\ref{hsol}) and  (\ref{hsollast}) results in
\bea
C_{2j+1} & = & \frac{(-1)^{2j+1}}{\epsilon_{2j-1} \det M_{(2j,2j)} }[\alpha_{2j-1}M_{2j,2j-3}-\beta_{2j-1}M_{2j,2j-2} \non \\
& + & (E+\gamma_{2j-1})M_{2j,2j-1}-\delta_{2j-1}M_{2j,2j}],
\eea
which is equivalent to to a determinant of a matrix with two identical rows, automatically giving the boundary condition
\be
\label{boundary1}
C_{2j+1} = 0.
\ee
It is possible to show that the boundary condition $ C_{2j+2}=0 $ is equivalent to the quantization condition (\ref{quantization}).
The recursion relation for $ k=2j $ is
\be
\label{generalrecursions}
0=\alpha_{2j}C_{2j-2}+\beta_{2j}C_{2j-1}+(E+\gamma_{2j})C_{2j}+\delta_{2j}C_{2j-1}+\epsilon_{2j}C_{2j+2},
\ee
and therefore $ C_{2j+2} $ is
\be
C_{2j+2} = \frac{-1}{\epsilon_{2j}}[\alpha_{2j}C_{2j-2}+\beta_{2j}C_{2j-1}+(E+\gamma_{2j})C_{2j}].
\ee
Substituting the coefficients $ C_k $ from equations (\ref{hsol}), (\ref{hsollast}) and (\ref{boundary1}) leads to
\bea
C_{2j+2} & = & \frac{-(-1)^{2j}}{\epsilon_{2j} \det M_{(2j,2j)} }[\alpha_{2j}M_{2j2j-2}-\beta_{2j}M_{2j,2j-1}+(E+\gamma_{2j})M_{2j,2j}] \non \\
& = & \frac{(-1)^{2j+1}}{\epsilon_{2j}} \; \frac{ \det {\bf M}_{2j+1}}{\det M_{(2j,2j)}} =  \frac{(-1)^{2j+1}}{\epsilon_{2j}} \; \frac{P_{2j+1}(E)}{\det M_{(2j,2j)}},
\eea
which means that the quantization condition coincides with the boundary condition $ C_{2j+2}=0 $ for the most general \qes potential.

These calculations are not valid if $ \det M_{(2j,2j)}=0 $ or if $ \eps_{k}=0 $, but for all other \qes potentials the wave function coefficients are given by the minors of the coefficient matrix $ {\bf M}_{(2j+1)} $ with the quantization condition coinciding with the boundary condition $ C_{2j+2}=0 $ and the boundary condition $ C_{2j+1}=0 $ automatically satisfied.
The special case of $ \eps_{k}=0 $ is discussed in section (\ref{epsilonzero}).

%%%%%%%%%%%%%%%%%%%%%%%%%%%%%%%%%%%%%%%%%%%%%%%%%%%%%%%%%%%%
\subsection{The Secular Polynomials}
%%%%%%%%%%%%%%%%%%%%%%%%%%%%%%%%%%%%%%%%%%%%%%%%%%%%%%%%%%%%

\label{secularp}
In this section $ j $ is kept as a free parameter.
The secular polynomials of the \qes problems which transform into matrices (\ref{matrix}), which are four-diagonal or less, may be calculated by a recursion relation.
In these cases the recursion for the secular polynomials is of the same degree as the recursion for the wave function coefficients (\ref{generalrecursions}).
This does not imply that the two sets of polynomials are of the same degree.
For example for $ \alpha_k = 0 $ the determinant of the matrix $ M_{k+1} $ as defined in equation (\ref{matrix}) can be calculated by
\bea
P_{k+1}(E) & = & (E+\gamma_k) P_k(E) - \beta_k \delta_{k-1} P_{k-1}(E) \non \\
& + & \beta_k \beta_{k-1} \delta_{k-2} P_{k-2}(E)
\eea
with initial conditions $ P_0=1 , \; P_1=\gamma_0, \; P_2=\gamma_0\gamma_1-\beta_1\delta_0 $.
$ P_k(E) $ is a polynomial of degree $ k $ whereas the coefficient $ C_k $ is not (the case where $ \eps_k = 0 $ is described in section (\ref{epsilonzero})).

For the most general potentials, which have a five diagonal coefficient matrix (\ref{matrix}), a new recursion relation for calculating the secular polynomials is derived.
This recursion relation has more than 5 terms as shown in the following calculation.  

We start with the coefficient matrix $ {\bf M}_{k+1} $
\be
{\bf M}_{k+1} = \left( \ba{cccccccc}
\gamma_0 & \delta_0 & \epsilon_0 & 0 & 0 & 0 & \ddots \\
\beta_1 & \gamma_1 & \delta_1 & \epsilon_1 & 0 & \ddots & 0 \\
\alpha_2 & \beta_2 & \gamma_2 & \delta_2 & \ddots & 0 & 0 \\
0 & \alpha_3 & \beta_3 & \ddots & \delta_{k-3} & \epsilon_{k-3} & 0 \\
0 & 0 & \ddots & \beta_{k-2} & \gamma_{k-2} & \delta_{k-2} & \epsilon_{k-2} \\
0 & \ddots & 0 & \alpha_{k-1} & \beta_{k-1} & \gamma_{k-1} & \delta_{k-1} \\
\ddots & 0 & 0 & 0 & \alpha_{k} & \beta_{k} & \gamma_{k}
\ea
\right).
\ee
Now let us define the matrix $ {\bf N}_k $ that is a specific minor of $ {\bf M}_{k+1} $ - the minor of the $ \beta_k $ term
\be
{\bf N}_k = \left( \ba{cccccccc}
\gamma_0 & \delta_0 & \epsilon_0 & 0 & 0 & 0 & \ddots \\
\beta_1 & \gamma_1 & \delta_1 & \epsilon_1 & 0 & \ddots & 0 \\
\alpha_2 & \beta_2 & \gamma_2 & \delta_2 & \ddots & 0 & 0 \\
0 & \alpha_3 & \beta_3 & \ddots & \delta_{k-4} & \epsilon_{k-4} & 0 \\
0 & 0 & \ddots & \beta_{k-3} & \gamma_{k-3} & \delta_{k-3} & 0 \\
0 & \ddots & 0 & \alpha_{k-2} & \beta_{k-2} & \gamma_{k-2} & \eps_{k-2} \\
\ddots & 0 & 0 & 0 & \alpha_{k-1} & \beta_{k-1} & \delta_{k-1}
\ea
\right).
\ee
The determinant of $ {\bf M}_{k+1} $ in terms of its minors of the forms $ {\bf M}_i $ and $ {\bf N}_i $ is 
\bea
\det({\bf M}_{k+1}) & = & (E+\gamma_k) \det({\bf M}_{k}) - \beta_k \det({\bf N}_{k}) \non \\
& + & \alpha_k \delta_{k-1} \det({\bf N}_{k-1}) - \alpha_k (E+\gamma_{k-1}) \eps_{k-2} \det({\bf M}_{k-2}) \non \\
& + & \alpha_k \alpha_{k-1} \eps_{k-2} \eps_{k-3} \det({\bf M}_{k-3})
\eea
whereas the determinant of $ {\bf N}_{k} $ in terms of its minors of the forms $ {\bf M}_i $ and $ {\bf N}_i $ is
\bea
\det({\bf N}_{k}) & = & \delta_{k-1} \det({\bf M}_{k-1}) -  \beta_{k-1} \eps_{k-2} \det({\bf M}_{k-2}) \non \\
& + & \alpha_{k-1} \eps_{k-2} \det({\bf N}_{k-2}).
\eea
The determinant of the matrix $ {\bf N}_{k} $ is a polynomial of degree $ k $ that is denoted by $ Q_k $
\be
Q_k = \det({\bf N}_{k}).
\ee
The determinant $ P_{k+1} $ is therefore
\bea
\label{5tpq1}
P_{k+1} & = & (E+\gamma_k) P_k - \beta_k Q_k + \alpha_k \delta_{k-1} Q_{k-1} \non \\
& - & \alpha_k (E+\gamma_{k-1}) \eps_{k-2} P_{k-2} + \alpha_k \alpha_{k-1} \eps_{k-2} \eps_{k-3} P_{k-3}
\eea
and the determinant $ Q_k $ is
\be
\label{5tpq2}
Q_k = \delta_{k-1} P_{k-1} -  \beta_{k-1} \eps_{k-2} P_{k-2} + \alpha_{k-1} \eps_{k-2} Q_{k-2}.
\ee
To obtain a recursion that will include only the secular polynomials $ P_k $ the polynomials $ Q_k, Q_{k-1}, Q_{k-2} $ must be replaced.
A third equation, equation (\ref{5tpq1}) with $ k \ra k-1 $ is added
\bea
\label{5tpq3}
P_{k} & = & (E+\gamma_{k-1}) P_{k-1} - \beta_{k-1} Q_{k-1} + \alpha_{k-1} \delta_{k-2} Q_{k-2} \non \\
& - & \alpha_{k-1} (E+\gamma_{k-2}) \eps_{k-3} P_{k-3} + \alpha_{k-1} \alpha_{k-2} \eps_{k-3} \eps_{k-4} P_{k-4}.
\eea
The equations (\ref{5tpq1}, \ref{5tpq2}, \ref{5tpq3}) form a set of three equations in the variables: $ Q_k, Q_{k-1}, Q_{k-2} $.
Solving these equations and substituting  $  Q_{k-1}, Q_{k-2} $ in equation (\ref{5tpq3}) gives the wanted recursion relation for $ P_{k}(E) $
\bea
0 & = & [-\alpha_{k-2}\delta_{k-4}\delta_{k-3}+\beta_{k-3}\beta_{k-2}\eps_{k-4}]P_{k} \non \\
& + & [\beta_{k-3}(-\beta_{k-2}(E+\gamma_{k-1})+\alpha_{k-1}\delta_{k-2}) \eps_{k-4} \non \\
& + & \alpha_{k-2}\delta_{k-4}((E+\gamma_{k-1})\delta_{k-3}-\beta_{k-1}\eps_{k-4})] P_{k-1} \non \\ 
& + & [\delta_{k-2} (\beta_{k-3} (\beta_{k-2}\beta_{k-1}-\alpha_{k-1} (E+\gamma_{k-2}))\eps_{k-4} \non \\
& + & \alpha_{k-2}\delta_{k-3} (-\beta_{k-1}\delta_{k-4}+\alpha_{k-1}\eps_{k-4})) \non \\
& + & \alpha_{k-2}\beta_{k-1} ((E+\gamma_{k-2})\delta_{k-4}-\beta_{k-2}\eps_{k-4}) \eps_{k-3} ] P_{k-2} \non \\
& + & [\alpha_{k-2}\alpha_{k-1}\delta_{k-4}\delta_{k-3}(\delta_{k-3}\delta_{k-2}-(E+\gamma_{k-3})\eps_{k-4}-(E+\gamma_{k-2})\eps_{k-3}) \non \\
& + & \beta_{k-2}(\alpha_{k-2}\beta_{k-1}(E+\gamma_{k-3})+ \beta_{k-3}(-\beta_{k-2}\beta_{k-1} \non \\
& + & \alpha_{k-1}(E+\gamma_{k-2})))\eps_{k-4}\eps_{k-3}] P_{k-3} \non \\
& + & [\alpha_{k-2}\alpha_{k-1}\beta_{k-3}(-\delta_{k-4}\delta_{k-3}\delta_{k-2}+(E+\gamma_{k-3})\delta_{k-2}\eps_{k-4} \non \\
& - & \beta_{k-2}\eps_{k-4}\eps_{k-3})\eps_{k-4}+\alpha_{k-2}\delta_{k-4}(\beta_{k-3}\beta_{k-2}\beta_{k-1}\non \\
& + & \alpha_{k-2}(-\beta_{k-1}(E+\gamma_{k-3})+\alpha_{k-1}\delta_{k-3}))\eps_{k-4}\eps_{k-3}] P_{k-4} \non \\
& + & [\alpha_{k-3}\alpha_{k-2}(\alpha_{k-1}(E+\gamma_{k-4})\delta_{k-3}\delta_{k-2}-\alpha_{k-1}\beta_{k-3}\delta_{k-2}\eps_{k-4}\non \\
& + & \beta_{k-1}(-\beta_{k-2}(E+\gamma_{k-4})+\alpha_{k-2}\delta_{k-4})\eps_{k-3})\eps_{k-5}\eps_{k-4}] P_{k-5} \non \\
& + & [\alpha_{k-4}\alpha_{k-3}\alpha_{k-2}(-\alpha_{k-1}\delta_{k-3}\delta_{k-2}\non \\
& + & \beta_{k-2}\beta_{k-1}\eps_{k-3})\eps_{k-6}\eps_{k-5}\eps_{k-4}] P_{k-6}.
\eea
Initial conditions for this set of equations are the first six polynomials $ P_0, \; P_1, \; P_2, \; P_3, \; P_4, \; P_5 $.
This is a 7 term recursion but it becomes a four (or less) term relation if one or more of the coefficients $ \alpha _k, \; \beta _k, \; \delta_k, \; \eps _k $ are chosen to be identically zero.

%%%%%%%%%%%%%%%%%%%%%%%%%%%%%%%%%%%%%%%%%%%%%%%%%%%%%%%%%%%%
\subsection{Parity of the \qes Sector Wave Functions}
%%%%%%%%%%%%%%%%%%%%%%%%%%%%%%%%%%%%%%%%%%%%%%%%%%%%%%%%%%%%

For every even potential (not only QES) the Hamiltonian can be separated into two blocks with different parities.
In the \qes potentials either one or both of these blocks are \qes. If $\xi(x)$ is an odd function both blocks are always \qes. On the other hand if $\xi(x)$ is an even function the generic case has only one \qes block.
In a certain family of \qes potentials with even $\xi(x)$ for some of the potentials the \qes block is the even one, and for others it is the odd one.
The difference is in the parity of $e^{a(x)}$. 

%%%%%%%%%%%%%%%%%%%%%%%%%%%%%%%%%%%%%%%%%%%%%%%%%%%%%%%%%%%%
\subsubsection{\qes Sectors of different parities}
%%%%%%%%%%%%%%%%%%%%%%%%%%%%%%%%%%%%%%%%%%%%%%%%%%%%%%%%%%%%

The general transformation from a potential with even parity wave functions to a potential of the same family (the same $\xi(x)$) with odd parity wave functions can be derived by changing the linear terms in the quasi-gauge Hamiltonian (\ref{qghamiltonian}).
A family of potentials with even $ \xi(x) $ and even $ e^{-a(x)} $ is considered.
What is the closest family with the same $ \xi(x) $ but odd $ e^{-a(x)} $ ?
The new Hamiltonian is connected to the original by
\bea
H^{odd}_G & = & H_G + \sum_{a= \pm ,0} \hat{c}_a T^a \\
& = & - \half p_4(\xi)\frac{d^2}{d \xi^2}+\left[p_3(\xi)+\hat{p}_3(\xi) \right] \frac{d}{d\xi}+\left[ p_2(\xi)+ \hat{p}_2 \right],
\eea
where $ \h{p}_3(\xi) $ and $ \h{p}_2(\xi) $ are the corrections to $ p _3(\xi) $ and $ p_2(\xi) $
\bea
\h{p}_3(\xi) & = & -\hat{c}_+\xi^2 +\hat{c}_0\xi+\hat{c}_- \\
\h{p}_2(\xi) & = & 2j\hat{c}_-\xi.
\eea
The odd wave function is 
\be
\psi_{odd}(x)=e^{-a_{odd}(x)} \sum _{k=0}^{2j}\hat{C}_k \xi^k(x).
\ee
The coefficients $ \hat{C}_k $ differ from the wave function coefficients $ C_k $ in the original potential but they are calculated using the same formalism.
The new relevant functions (\ref{aderivative}-\ref{deltav}) are
\bea
\label{oddaderivative}
{\cal A}_{odd} (x) & = & \frac{p'_4/4 + p_3 + \hat{p}_3}{\sqrt{p_4}} = \left. {\cal A}(x) + \frac{\hat{p}_3}{\sqrt{p_4}} \right|_{\xi=\xi(x)}\\
\label{oddaphase}
a_{odd}(x) & = &  \int \frac {p'_4/4 + p_3 + \hat{p}_3}{p_4}d\xi = a(x) + \int^{\xi(x)} \frac {\hat{p}_3}{p_4}d\xi \\
\label{odddeltav}
\Delta V_{odd}(x) & = & p_2(\xi) + \hat{p}_2(\xi) = \left. \Delta V (x) + \hat{p}_2(\xi) \right|_{\xi=\xi(x)}
\eea
The wave functions of the \qes sector have the explicit form 
\be
\label{oddpotential}
\psi_{odd}(x) = \exp \left\{- \int ^{\xi(x)} \frac {\hat{p}_3}{p_4}d\xi \right\} e^{-a(x)} \sum _{k=0}^{2j} \hat{C}_k \xi^k(x).
\ee
The function $ \exp \left\{- \int ^{\xi(x)} \frac {\hat{p}_3}{p_4}d\xi \right\} $ is chosen to be odd.
As in (\ref{potential}) the corresponding potential is
\bea
V_{odd}(x) & = & \Delta V_{odd}(x) + \half {\cal A}_{odd}^2(x) - \half {\cal A}'_{odd}(x) \\
& = & V(x) + \left[ \hat{p}_2 + {\cal A} \frac{\hat{p}_3}{\sqrt{p_4}} - \half \hat{p}'_3 + \half \frac { \hat{p}_3}{p_4} \left( \hat{p}_3 + \half p'_4 \right) \right]_{\xi=\xi(x)}
\eea
The potential $ V_{odd}(x) $ is the most similar to $ V(x) $ that has an odd \qes sector.

%%%%%%%%%%%%%%%%%%%%%%%%%%%%%%%%%%%%%%%%%%%%%%%%%%%%%%%%%%%%
\subsection{Example 2 - Potential Built From Hyperbolic Functions - Sine}
%%%%%%%%%%%%%%%%%%%%%%%%%%%%%%%%%%%%%%%%%%%%%%%%%%%%%%%%%%%%

The hyperbolic sine potential is an example that demonstrates many of the things discussed earlier.
The \qg Hamiltonian in this example is
\bea
H_G & = & - \half \left( T^0T^0-T^-T^- \right) -jT^0-a \left(T^-+T^+ \right) \\
& = & - \half (\xi^2-1)d^2_\xi + \left( a \xi^2 - a - \frac{\xi}{2} \right) d_\xi - 2ja \xi,
\eea
and the relevant functions (\ref{xi}-\ref{deltav}) are
\bea
\xi (x) & = & \cosh x \\
{\cal A}(x) & = & a \sinh x \\
a(x) & = & a \cosh x \\
\Delta V(x) & = & -2ja \cosh x.
\eea
Equation (\ref{potential}) gives the hyperbolic potential
\be
V(x)= \frac{a^2}{2} \sinh ^2x-a \left( 2j+ \frac{1}{2} \right) \cosh x.
\ee
\begin{figure}[h]
\epsfig{file=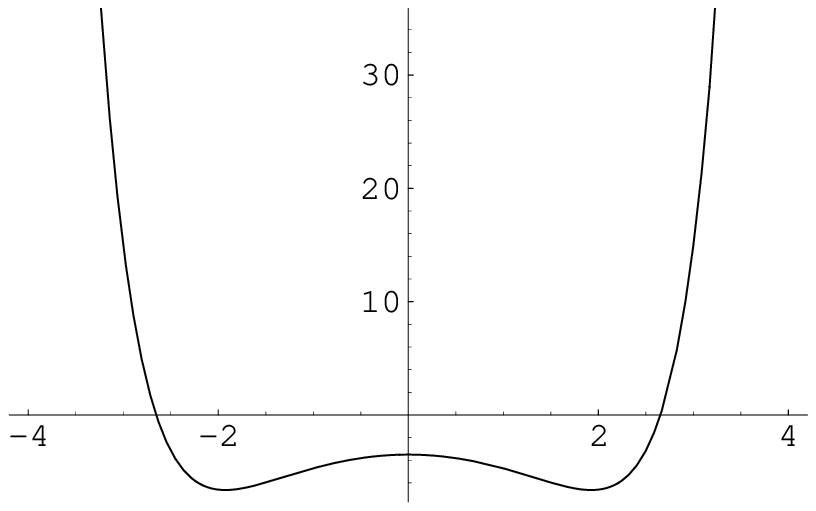}
\caption{The hyperbolic sine potential with $ a=1 $, $ j = \frac{3}{2} $.}
\end{figure}

The wave function is
\be
\label{wavesine}
\psi = e^{-a \cosh x}\sum_{k=0}^{2j}C_k (\cosh x)^k.
\ee
The recursion relation that corresponds to the \schr equation is
\bea
0 & = & a(2j+1-k)C_{k-1}+ \left( E+ \frac{1}{2}k^2 \right)C_k \nonumber \\ & + & a(k+1)C_{k+1}-\frac{1}{2}(k+1)(k+2) C_{k+2},
\eea
with the boundary conditions $ C_{-1}=0 , \; C_{2j+1}=0 , \; C_{2j+2}=0 $.
The recursion relation for the secular polynomials is
\bea
P_{k+1} & = & \left[E+ \frac{1}{2}k^2 \right] P_k - a^2k(2j+1-k)P_{k-1} \nonumber \\
& - & \half a^2k(k-1)(2j+1-k)(2j+2-k)P_{k-2},
\eea
with the initial conditions $ P_0=1 , \; P_1=E , \; P_2=E^2+\half E - 2j \: a^2 $.

The odd potential that is closest to this potential is obtained by transforming
\be
H_G \ra H_G-T^0.
\ee
Under this transformation $ \xi(x) $ is unchanged and the relevant functions (\ref{oddaderivative}-\ref{odddeltav}) are
\bea
{\cal A}_{odd}(x) & = & a \sinh x - \coth x\\
a(x)_{odd} & = & a \cosh x - \ln ( \sinh x ) \\
\Delta V(x)_{odd} & = & -2ja \cosh x.
\eea
The odd parity \qes sector potential (\ref{oddpotential}) is
\be
V_{odd}(x)= \frac{a^2}{2} \sinh ^2x - a \left( 2j+ \frac{3}{2} \right) \cosh x + \half.
\ee

The odd wave function is
\be
\psi = \sinh x e^{-a \cosh x}\sum_{k=0}^{2j}\h{C}_k (\cosh x)^k
\ee
The recursion relation for the coefficients is
\bea
0 & = & a(2j+1-k)\h{C}_{k-1}+ \left[ E+ \frac{1}{2}k(k+2) \right] \h{C}_k \nonumber \\
& + & a(k+1)\h{C}_{k+1}-\frac{1}{2}(k+1)(k+2) \h{C}_{k+2},
\eea
with the boundary values $ \h{C}_{-1}=0 , \; \h{C}_{2j+1}=0 , \; \h{C}_{2j+2}=0 $.
The recursion relation for the secular polynomials is
\bea
\h{P}_{k+1} & = & \left[E+ \frac{1}{2}k(k+2) \right] \h{P}_k - a^2k(2j+1-k)\h{P}_{k-1} \nonumber \\
& - & \half a^2k(k-1)(2j+1-k)(2j+2-k)\h{P}_{k-2}
\eea
with the initial conditions $ \h{P}_0=1 , \; \h{P}_1=E , \; \h{P}_2=E^2 +E -2j \: a^2 $.
The wave function is the same as in (\ref{wavesine}) but the coefficients $ C_k $ are different.

%%%%%%%%%%%%%%%%%%%%%%%%%%%%%%%%%%%%%%%%%%%%%%%%%%%%%%%%%%%%
\section{Special Case - $ \epsilon _k=0 $}
%%%%%%%%%%%%%%%%%%%%%%%%%%%%%%%%%%%%%%%%%%%%%%%%%%%%%%%%%%%%

\label{epsilonzero}
In all the cases with $ \epsilon _k=0 $ ($ \eps_k $ is defined in equation (\ref{abcde}), $ \epsilon _k=0 $  is  equivalent to $ c_{--}=0 $ in $ H_G $), $ C_k $ can be polynomials of degree $ k $ if the free parameter is chosen to be $ C_0=1 $.
In this section it is proved that $ P_k $ and $ C_k $ are identical up to an overall proportion factor $ \Delta_k $.
This means that the initial boundary solution of the \schr equation (for a general discussion see Appendix \ref{initialboundary}) is the generating function of the secular polynomials.
This property was shown in \cite{benderdunne96,finkel96} to hold for specific examples, and is here expanded to include the general group of potentials with $ \epsilon _k=0 $.

If $ \eps_k = 0 $, then the recursion relation for the wave function coefficients is
\be
\label{easy4recursion}
0=\alpha_kC_{k-2}+\beta_kC_{k-1}+(E+\gamma_k)C_k+\delta_kC_{k+1}.
\ee
Since this is a four term recursion relation, only 3 boundary value conditions are necessary
\be
\label{boundary4term}
C_{-2}=0 , \; C_{-1}=0 , \; C_{2j+1}=0.
\ee
The recursion relation for the secular polynomials, derived independently from the matrix (\ref{matrix}) using the known properties of determinants is
\bea
\label{easy4polynomials}
P_{k+1}(E) & = & (E+\gamma_k)P_k(E)-\beta_k\delta_{k-1}P_{k-1}(E) \non \\
& + & \alpha_k\delta_{k-1}\delta_{k-2}P_{k-2},
\eea
with the initial value conditions
\be
\label{initial4term}
P_0=1 , \; P_1=\gamma_0 , \; P_2=\gamma_0\gamma_1-\beta_1\delta_0.
\ee
From the two recursion relations a connection between the wave function coefficients and the polynomials can be derived:
\be
\label{easy4coef}
C_k= (-1)^k \left( \prod_{i=0}^{k-1}\frac{1}{\delta_i} \right) P_k(E) = \Delta_k P_k(E)
\ee
The quantization condition $ P_{2j+1}(E) = 0 $ coincides with the boundary condition $ C_{2j+1} = 0 $.
The wave function can now be written as
\be
\psi_n(x) = e^{-a(x)} \sum_{k=0}^{2j} \Delta_k P_k(E_n) \xi^k(x)
\ee
where $ E_n $ are the eigenvalues of the \schr equation (\ref{easy4coef}) into equation (\ref{generalrecursions}) but leaving the value of $ E $ undetermined gives a generating function for the secular polynomials of this system.
The function
\be
\label{generating4term}
\Psi(x) = e^{-a(x)} \sum_{k=0}^\infty \Delta_k P_k(E) \xi^k(x)
\ee
where $ E $ is a free parameter, is the initial value solution of the \schr equation (\ref{schr}).
It is also the generating function of the secular polynomials $ P_k(E) $.

%%%%%%%%%%%%%%%%%%%%%%%%%%%%%%%%%%%%%%%%%%%%%%%%%%%%%%%%%%%%
\subsection{Example 3 - Potential Built From Hyperbolic Functions - Cosecant}
%%%%%%%%%%%%%%%%%%%%%%%%%%%%%%%%%%%%%%%%%%%%%%%%%%%%%%%%%%%%

\begin{figure}[h]
\epsfig{file=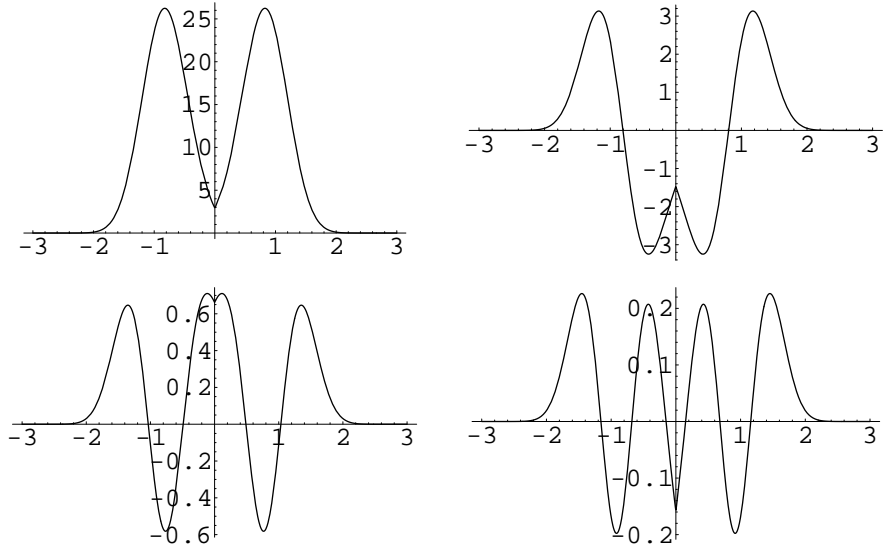}
\caption{The wave functions of the hyperbolic cosecant potential with $ j = \frac{3}{2} $.}
\label{newpwf}
\end{figure}

If the polynomial $ p_4(\xi) $ (\ref{hgp}) has a double root in $ \xi(x=0) $ then the function $ \xi(x) $ is defined separately for the different sections of the real line.
The Hamiltonian, the potential and the wave functions are chosen to have a well defined parity, either even or odd.
\be
V_-(-x) = \pm V_+(x), \; x>0.
\ee
The hyperbolic cosecant potential demonstrates this special quality discussed in \cite{normalizability93}.
It is generated by the quasi-gauge Hamiltonian
\bea
H_G & = & - \half (T^+ T^+ + T^0 T^0) + 2j (T^+ + T^-) - j T^0 \non \\
& = & \xi^2 (\xi^2 + 1)\frac{d^2}{d \xi ^2} + [(2j-1)\xi^3 - 2j \xi^2 - \half \xi^2  + 2j ] \frac{d}{d \xi} \non \\
& - & j(2j-1)\xi^2 + 4j^2\xi.
\eea
The relevant functions (\ref{xi}-\ref{deltav}) are
\bea
\xi(x) & = & - \frac{1}{\sinh x} = - \csch x \\
{\cal A}(x)  & = & - \frac{2j(\csch ^3 x + \csch ^2 - 1)}{\coth x \; \csch x} \\
a(x) & = & 2j (-2x + \sinh x + \ln \cosh x -\ln \sinh x) \\
\Delta V(x) & = & - j(2j-1)\xi^2 + 4j^2\xi,
\eea
and the potential (\ref{potential}) is
\bea
\label{cosecantv}
V(x) & = & -\frac{j}{4} \sech ^2 x (4-11j+12j \cosh 2x -j \cosh 4x \non \\
& + & (9+32j)\sinh x + \sinh 3x).
\eea
\begin{figure}[h]
\epsfig{file=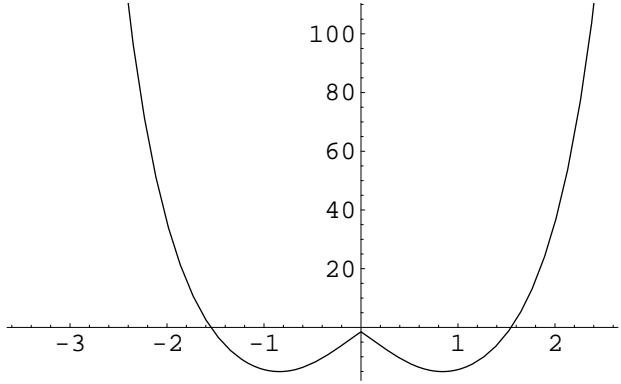}
\caption{The hyperbolic cosecant potential with $ j = \frac{3}{2} $.}
\end{figure}
In accordance with \cite{normalizability93} the potential and wave functions are defined to be even by replacing $ x $ with $ |x| $.
The recursion relation connecting the wave function coefficients is
\bea
0 & = & \half(2j+1-k)(2j+2-k)C_{k-2} - 2j(2j+1-k)C_{k-1} \non \\
& + & [E + \frac{k^2}{2}]C_k - 2j(k+1)C_{k+1}
\eea
with the boundary conditions $ C_{-2}=0,\;C_{-1}=0,\;C_{2j+1}=0 $. 
The secular polynomials are also connected by a four term recursion given by
\bea
P_{k+1}(E) & = & [E + \frac{k^2}{2}]P_k(E) - 4j^2k(2j+1-k)P_{k-1}(E) \non \\
& + & 2j^2(2j+1-k)(2j+2-k)k(k-1)P_{k-2}(E),
\eea
with the initial conditions $ P_0 = 1,\;P_1 = E,\;P_2 = E^2+\half E + 8j^3 $.
The first polynomials and their zeros are given in table (\ref{cosecant}).
\begin{table}[h]
\begin{tabular}{|l||l|l|} \hline
$ j $ & $ P_{2j+1}(E) $ & $ E $ \\ \hline \hline
$ 0 $ & $ E $ & $ 0 $ \\ \hline
$ \half $ & $ E^2 + \half E-1 $ & $ -1.28 , 0.781 $ \\ \hline
$ 1 $ & $ E^3 + \frac{5}{2}E^2 - 15E -8 $ & $ -5.12 , -0.5 , 3.12 $ \\ \hline
$ \frac{3}{2} $ & $ E^4 + 7E^3 - \frac{311}{4}E^2 - \frac{477}{2}E + 729 $ & $ -11.48 , -4.5 , 2.03 , 6.95 $ \\ \hline
$ 2 $ & $ \ba{l} E^5 + 15E^4 - \frac{1007}{4}E^3 - \frac{4659}{2}E^2 \\ + 13412E + 35072 \ea $ & $ \ba{c} -20.36 , -11.02 , -2.06 , \\ 6.20 , 12.24 \ea $ \\ \hline
\end{tabular}
\caption{\label{cosecant} The \qes sector energy levels of the hyperbolic cosecant potential.}
\end{table}
The relation between the coefficients and the secular polynomials is
\be
C_k = \frac{1}{(2j)^k k!}P_k(E).
\ee
The wave functions have the form
\be
\psi_n = e^{- 2j (\sinh x - 2x)} (\tanh x)^{2j} \sum_{k=0}^{2j}\frac{(2j)^k}{k!}P_k(E_n)(-\csch x)^k
\ee
where $ E_n $ is the eigenvalue of the \schr equation.
The 4 analytical wave functions of the potential (\ref{cosecantv}) with $ j=3/2 $ are illustrated in figure (\ref{newpwf}).
The generating function of the secular polynomials is given by (\ref{generating4term})
\be
\Psi(x) = e^{-2j(\sinh x -2x)}(\tanh x)^{2j} \sum_{k=0}^\infty \frac{1}{(2j)^k k!} P_k(E)(-\csch x)^k
\ee
where $ j $ is a free parameter of the potential.

%%%%%%%%%%%%%%%%%%%%%%%%%%%%%%%%%%%%%%%%%%%%%%%%%%%%%%%%%%%%
\subsection{Example 4 - Polynomial Potential}
%%%%%%%%%%%%%%%%%%%%%%%%%%%%%%%%%%%%%%%%%%%%%%%%%%%%%%%%%%%%

\label{poly}
The potential discussed in the introduction is one of a family of polynomial potentials that have the \qg Hamiltonian
\bea
H_G & = & -2T^0T^- - (2j+1)T^- - \nu T^+ + \mu T^0 \\
& = & - 2 \xi  \frac{d^2}{d \xi ^2} + ( \nu \xi ^2 + \mu \xi -1) \frac{d}{d \xi} - 2j \nu \xi,
\eea
and the relevant functions (\ref{xi}-\ref{deltav})
\bea
\xi (x) & = & x^2 \\
{\cal A}(x)  & = & \half (\nu x^3 + \mu x) \\
a(x) & = & \frac{\nu x^4}{8} + \frac {\mu x^2}{4} \\
\Delta V(x) & = & - 2j \nu x^2.
\eea
The potential (\ref{potential}) is
\be
\label{polypotential}
V(x) = \frac{\nu^2}{8}x^6+\frac{\mu \nu}{4}x^4+\left[\frac{\mu^2}{8}-2\nu\left(j+\frac{3}{8}\right)\right]x^2
\ee
(where the sextic potential is a polynomial potential with $ \nu = 2,\; \; \mu =0 $).
The general recursion relation for the coefficients of the wave function is
\be
0 = \nu(2j+1-k)C_{k-1} + (E - \mu k)C_k + (2k+1)(k+1)C_{k+1}
\ee
with boundary value conditions $ C_{-1}=0 , \; C_{2j+1}=0 $. 
The secular polynomials satisfy the recursion relation
\be
P_{k+1} = (E-\mu k)P_k - \nu(2j+1-k)(2k-1)kP_{k-1}
\ee
with initial conditions $ P_0(E)=1 , \; P_1(E)=E $.
The generating function of the secular polynomials is
\be
\label{generatingpoly}
\Psi(x) = e^{- \left( \frac{\nu x^4}{8} + \frac {\mu x^2}{4} \right)} \sum_{k=0}^\infty \frac{(-2)^k}{(2k)!} P_k(E) x^{2k}.
\ee

%%%%%%%%%%%%%%%%%%%%%%%%%%%%%%%%%%%%%%%%%%%%%%%%%%%%%%%%%%%%
\subsection{Example 5 - Exponential Potential}
%%%%%%%%%%%%%%%%%%%%%%%%%%%%%%%%%%%%%%%%%%%%%%%%%%%%%%%%%%%%

The exponential potential has no well defined parity because $ \xi(x) $ has no zeros \cite{normalizability93}.
The \qg Hamiltonian for generating this potential is
\bea
H_G & = & -\frac{1}{2}T^0T^0-aT^++cT^-+\left( b-j+\frac{1}{2}\right) T^0 \\
& = & -\frac{1}{2}\xi^2d_{\xi}^2+(a\xi^2+b\xi+c)d_{\xi}-2aj\xi,
\eea
and the relevant functions (\ref{xi}-\ref{deltav}) are
\bea
\xi (x) & = & e^x \\
{\cal A} (x) & = & ae^x+\left(b+\frac{1}{2}\right)+ce^{-x} \\
a(x) & = & ae^x+\left(b+\frac{1}{2}\right)x-ce^{-x} \\
\Delta V(x) & = & - 2 a j e^x.
\eea
The potential (\ref{potential}) is
\be
V(x)=\frac{1}{2}[ae^x+(b-2j)]^2+\frac{1}{2}[ce^{-x}+(b+1)]^2.
\ee
The wave function coefficients recursion relation is
\be
0= a(2j+1-k)C_{k-1}+\left[E+k\left( \frac{k}{2} - \half -b \right)\right]C_k - c(k+1)C_{k+1}
\ee
with the boundary conditions $ C_{-1}=0 , \; C_{2j+1}=0 $. 
The recursion relation for the secular polynomials is
\bea
P_{k+1}(E) & = & \left[E+k\left(\frac{k}{2}- \half -b \right)\right] P_k(E) \non \\
& + & ack(2j+1-k)P_{k-1}(E)
\eea
with the initial value conditions $ P_0(E)=1 , \; P_1(E)=E $.
The relation between $ C_k $ and $ P_k(E) $ is
\be
C_k=\frac{1}{k!c^k}P_k(E).
\ee

%%%%%%%%%%%%%%%%%%%%%%%%%%%%%%%%%%%%%%%%%%%%%%%%%%%%%%%%%%%%
\section{Special Case - Exactly Solvable Potentials}
%%%%%%%%%%%%%%%%%%%%%%%%%%%%%%%%%%%%%%%%%%%%%%%%%%%%%%%%%%%%

\label{exactlysolvable}
The exactly solvable potentials can be described as \qe solvable potentials with an infinite \qe solvable sector \cite{shifmanreview,bagchi03}. 
This requires that the \qg Hamiltonian $ H_G $ does not depend on the dimension of the representation $ j $. 
The general \qg Hamiltonian that is independent of $ j $ is
\be
H_G = -\half Q_2(\xi) \frac{d^2}{d\xi^2} + Q_1(\xi) \frac{d}{d\xi},
\ee
where $ Q_2(\xi) $ is a polynomial of the second degree or less
\be
 Q_2(\xi) = -2c_{00}\xi^2 - 4c_{0-}\xi - 2c_{--},
\ee
and $ Q_1(\xi) $ is a polynomial of the first degree or less
\be
 Q_1(\xi) = (c_0-(2j-1)c_{00})\xi + (c_--(2j-1)c_{0-}).
\ee
$\alpha_k $ and $\beta_k$ in (\ref{abcde}) are zero for all exactly solvable potentials.
The coefficient matrix (\ref{matrix}) is therefore top triangular, which means that its determinant is equal to the infinite polynomial
\be
P(E) = \prod_{k=0}^{\infty} \gamma_k.
\ee
All of the wave functions can be analytically calculated and are of the form
\be
\psi_n = e^{-a(x)} \sum_{k=0}^{n-1} C_k(E_n)\xi^k(x).
\ee
Unlike the \qes potentials, the $ n $th  wave function is a polynomial of degree $ n $.
The coefficients $ C_k $ are given by the recursion relation
\be
0=(E+\gamma_{2j})C_{2j}+\delta_{2j}C_{2j-1}+\epsilon_{2j}C_{2j+2}
\ee
For $ \eps_k = 0 $ or $ \delta_k = 0 $ all of the coefficients are known in closed form.
The best known examples of exactly solvable potentials are given in table (\ref{exacttable}). 
More examples can be created by the factorization method known as Infeld-Hull \cite{infeldhull,hermann81}.
\begin{table}[h]
\begin{tabular}{|l|l|} \hline
$V(x)$ & $H_G$ \\ \hline \hline
Harmonic Oscillator $ \frac{\omega^2}{2}x^2 $ & $ -\half T^- T^- + \omega T^0 $ \\ \hline
Harmonic Oscillator $ \half x^2 -\half $ & $ -2T^0T^- - (2j+1)T^- + 2T^0 $ \\ \hline
Morse Potential $ A(e^{-2\alpha x}-2e^{-\alpha x}) $ & $ \ba{l} -\frac{\alpha^2}{2} T^0 T^0 + \sqrt{2A} \alpha T^- \\- \alpha(\sqrt{2A}+(j+\half)\alpha)T^0 \ea $ \\ \hline
P\"oschl-Teller Potential $- \frac{U_0}{\cosh^2\alpha x}$ & $ \ba{l} -\frac{\alpha^2}{2}[T^0 T^0 + T^- T^- \\ + (\sqrt{1+8\frac{8U_0}{\alpha^2}}-(2j+1))T^0] \ea $ \\ \hline
\end{tabular}
\caption{\label{exacttable}The exactly solvable potentials and their $ sl(2) $ representations.}
\end{table}

%%%%%%%%%%%%%%%%%%%%%%%%%%%%%%%%%%%%%%%%%%%%%%%%%%%%%%%%%%%%
\subsection{Example 6 - Harmonic Oscillator}
%%%%%%%%%%%%%%%%%%%%%%%%%%%%%%%%%%%%%%%%%%%%%%%%%%%%%%%%%%%%

\label{hoe}
The Harmonic oscillator can be described by two different \qg Hamiltonians. The first, $ -\half T^- T^- + \omega T^0 $ , gives both odd and even wave functions and their energies. 
The alternative representation separates the problem into even and odd parts but has the initial value solution that is used in chapter (\ref{approximations}).
The \qg Hamiltonian of the even states of the harmonic oscillator is given by
\bea
\label{hoqgham}
H_G & = & -2T^0T^- - (2j+1)T^- + 2T^0 \non \\
    & = & -2\xi \frac{d^2}{d\xi^2} + (2\xi-1) \frac{d}{d\xi}
\eea
with the relevant functions (\ref{xi}-\ref{aphase})
\bea
x(\xi) & = & \int \frac{d\xi}{\sqrt{4\xi}} = \sqrt \xi \non \\
{\cal A}(x) & = & \sqrt \xi = x \non \\
a(x) & = & \int^{\xi(x)} \frac{2\xi}{4\xi} d\xi = \half \xi = \half x^2 \non \\
\Delta V(x) & = & 0.
\eea
The potential (\ref{potential}) is
\be
V(x) = \half x^2 - \half.
\ee
The \qg Hamiltonian (\ref{hoqgham}) gives the even wave functions. 
The odd wave functions are given by a correction of $ -2T^- $ to (\ref{hoqgham}) that has odd $ e^{-a(x)} $ but the same potential $ V(x) $.  
The even wave functions are
\be
\psi_n^{even} = e^{-\half x^2} \sum_{k=0}^n C_k(E_n^{even}) x^{2k}
\ee
The wave functions coefficients $ C_k $ are given by a two-term recursion relation 
\be
0 = (E-2k)C_k + (2k+1)(k+1)C_{k+1}
\ee
with the boundary condition $ C_{-1} = 0 $.
The closed form of the coefficients $ C_k $ can be calculated
\be
C_k(E)=\frac{(-2)^k}{(2k)!}\prod_{n=0}^{k-1} (E-2n).
\ee
The even energy levels that correspond to these solutions are
\be
E_n^{even}=2n.
\ee
The wave functions for the even states have the closed form
\be
\psi_{2j}^{even} = e^{-\half x^2} \sum_{k=0}^{2j} \frac{(-2)^k}{(2k)!} \left[ \prod_{n=0}^{k-1} (E_{2j}^{even}-2n) \right] x^{2k}.
\ee
The polynomials $ P_{2j+1}(E) $ are
\be
P_{2j+1}^{even}(E) = \prod_{k=0}^{2j} (E-2k),
\ee
and the even wave functions are then written as
\be
\psi_{2j}^{even} = e^{-\half x^2} \sum_{k=0}^{2j} \frac{(-2)^k}{(2k)!} P_k^{even}(E_{2j}) x^{2k}.
\ee
The initial value solution of the \schr equation is the generating function of the polynomials $ P_{2j+1}^{even}(E) $
 \be
\label{generatingorigin}
\Psi^{even} = e^{-\half x^2} \sum_{k=0}^{\infty} \frac{(-2)^k}{(2k)!} P_k^{even}(E_) x^{2k}.
\ee
Now $ V(x) $ and $ E $ can be redefined to be
\be
V(x)=\half x^2, \; E_n^{even} = 2n + \half,
\ee
The form of $ \Psi^{even} $ is unchanged.

%%%%%%%%%%%%%%%%%%%%%%%%%%%%%%%%%%%%%%%%%%%%%%%%%%%%%%%%%%%%
\section{Special Case - A Complete Separation of Even and Odd States}
%%%%%%%%%%%%%%%%%%%%%%%%%%%%%%%%%%%%%%%%%%%%%%%%%%%%%%%%%%%%

A system which can be separated into even and odd parts is obtained from (\ref{hset}) by taking $ \beta_k=0 , \; \delta_k=0 $. 
The recursion is a 3 term relation of the form:
\be
\label{special3term}
0= \alpha_kC_{k-2}+(E + \gamma_k)C_k + \epsilon_kC_{k+2}
\ee
with boundary values
\be
\label{special3termcond}
C_{-2}=0 , \; C_{-1}=0 , \; C_{2j+1}=0, \; C_{2j+2}=0.
\ee
From a practical point of view, since $ \xi(x) $ is an odd function the total wave function can be written as a sum of an even function and an odd function with independent recursion relations
\be
\psi = \psi^{even} + \psi^{odd}.
\ee
The secular polynomial of the whole system is a product of the quantization conditions of the separate systems
\bea
P_{2j+1}(E) & = & \prod_{i=0}^{2j}(E-E_i) = \prod_{i=0}^{[j]}(E-E_i^{even})\prod_{i=0}^{[j-\half]}(E-E_i^{odd}) \non \\
& = & Q_{[j+1]}(E) q_{[j+\half]}(E).
\eea
The wave function is
\bea
\psi_{2j+1}(x) & = & \psi_{2j+1}^{even}(x) + \psi_{2j+1}^{odd}(x) \\
& = & e^{-a(x)}\sum_{0}^{[j]}\frac{Q_k(E)}{(2k)!}\xi^{2k}(x) + e^{-a(x)}\sum_{0}^{[j-\half]}\frac{q_k(E)}{(2k+1)!}\xi^{2k+1}(x) \non
\eea

%%%%%%%%%%%%%%%%%%%%%%%%%%%%%%%%%%%%%%%%%%%%%%%%%%%%%%%%%%%%
\subsubsection{Even States}
%%%%%%%%%%%%%%%%%%%%%%%%%%%%%%%%%%%%%%%%%%%%%%%%%%%%%%%%%%%%

Let us define a new integer variable $ m=0,1,...,[j] ; \; \; k=2m $.
The even wave function coefficients $ D_m = C_{2m} $ satisfy the recursion relations
\be
\label{even3term}
0 = \alpha_{2m} D_{m-1} + (E+ \gamma_{2m}) D_m + \epsilon_{2m} D_{m+1},
\ee
with the boundary conditions $ D_{-1}=0 , \; D_{2j+1}=0 $.
The even energy polynomials $ Q_m^j $ satisfy the recursion relation
\be
Q_{m+1}^j(E) = (E+ \gamma_{2m})Q_m^j(E) - \alpha_{2m} \epsilon_{2m-2}Q_{m-1}^j(E)
\ee
with the initial conditions $ Q_0=1 , \; Q_1=E $.
This problem is solved by the same rules described in section (\ref{epsilonzero}).
The relation between the coefficients and the polynomials is again a simple proportion relation
\be
D_m= (-1)^m Q_m(E) \prod_{i=0}^{m-1} \frac{1}{ \epsilon_{2i}}
\ee
The quantization condition for the even states is
\be
Q_{[j+1]}(E)=0.
\ee

%%%%%%%%%%%%%%%%%%%%%%%%%%%%%%%%%%%%%%%%%%%%%%%%%%%%%%%%%%%%
\subsubsection{Odd States}
%%%%%%%%%%%%%%%%%%%%%%%%%%%%%%%%%%%%%%%%%%%%%%%%%%%%%%%%%%%%

The integer $ m $ is redefined to be $ m=0,1,...,[j- \half ] ; \; \;  k=2m+1 $.
The odd coefficients $ d_m = C_{2m+1} $ satisfy the recursion relations
\be
\label{odd3term}
0 = \alpha_{2m+1} d_{m-1} + (E+ \gamma_{2m+1}) d_m + \epsilon_{2m+1} d_{m+1},
\ee
with the boundary conditions $ d_{-1}=0 , \; d_{2j+1}=0 $.
The odd energy polynomials $ q_m $ satisfy the recursion relation
\be
q_{m+1}(E) = (E+ \gamma_{2m+1})q_m(E) - \alpha_{2m+1} \epsilon_{2m-1} q_{m-1}(E)
\ee
with the initial conditions $ q_0=1 , \; q_1=E+\gamma_1 $.
The relation between the coefficients and the polynomials is once more a simple proportion relation
\be
d_m= (-1)^m \prod_{i=0}^{m-1} \frac{1}{ \epsilon_{2i+1}}q_m(E).
\ee
The quantization condition for the odd states is
\be
q_{[j+\half]}(E)=0.
\ee

%%%%%%%%%%%%%%%%%%%%%%%%%%%%%%%%%%%%%%%%%%%%%%%%%%%%%%%%%%%%
\subsection{Example 7 - Potential Built From Hyperbolic Functions - Cosine}
%%%%%%%%%%%%%%%%%%%%%%%%%%%%%%%%%%%%%%%%%%%%%%%%%%%%%%%%%%%%

The hyperbolic cosine potential is an example of complete separation of even and odd states.
It is generated by the \qg Hamiltonian
\bea
H_G & = & -\half T^+T^+ + T^0T^0 -\half T^-T^- + aT^0 \\
& = & - \half (1- \xi^2)^2 d^2_\xi + \left[ a\xi-(2j-1)\xi(1-\xi^2) \right] d_\xi - j(2j-1)\xi^2 \non,
\eea
with the relevant functions (\ref{xi}-\ref{deltav})
\bea
\xi (x) & = & \tanh x \\
{\cal A}(x) & = & (a \cosh ^2 x -2j) \tanh x \\
a(x) & = & \frac{a}{2} \cosh ^2 x -2j \ln (\cosh x) \\
\Delta V(x) & = & - j(2j-1) \tanh ^2 x.
\eea
The potential (\ref{potential}) is
\be
V(x)= \frac{a^2}{2} \cosh ^4x- \frac{a}{2} \left( a+4j+2 \right) \cosh^2 x. 
\ee
\begin{figure}[h]
\epsfig{file=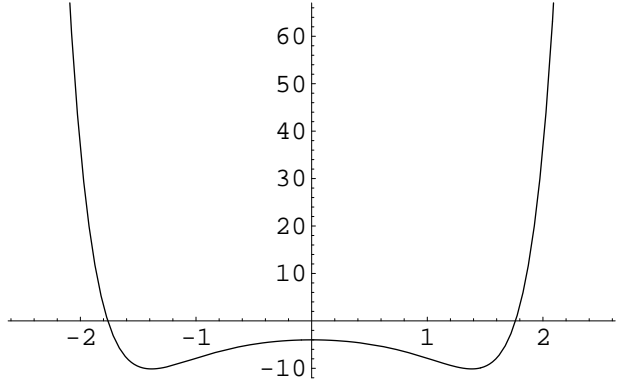}
\caption{The hyperbolic cosine potential with $ a=1 $, $ j = \frac{3}{2} $.}
\end{figure}
The recursion relation for the coefficients has the form of (\ref{special3term})
\bea
0 & = & \half(2j+2-k)(2j+1-k) C_{k-2} \nonumber\\
 & + & \left[E+k(2j-k-a) \right] C_k + \half(k+1)(k+2) C_{k+2}.
\eea
The boundary value conditions are given by $ C_{-2}=0,\;C_{-1}=0,\;C_{2j+1}=0,\;C_{2j+2}=0 $.

The recursion relation (\ref{even3term}) for the even state coefficients is
\bea
0 & = & (j+1-m)(2j+1-2m)D_{m-1} \nonumber \\
& + &[E+2m(2j-2m-a)]D_m \non \\
& + &(2m+1)(m+1)D_{m+1},
\eea
with boundary conditions $ D_{-1}=0 , \; D_{[j+1]}=0 $.
The recursion for the even energy polynomials $ Q_{m} $ is
\bea
Q_{m+1} & = & [E+ 2m(2j-a-2m)]Q_m \nonumber \\
& - & (j+1-m)(2j+1-2m)(2m-1)m Q_{m-1},
\eea
with initial conditions $ Q_0=1 , \; Q_1=E $.
The relation between the coefficients and the polynomials is
\be
D_m= \frac{(-2)^m}{(2m)!}Q_m(E)
\ee
The even wave function is
\be
\psi_n^{even} = e^{-a(x)}\sum_{0}^{[j]}\frac{Q_k(E_n^{even})}{(2k)!}\xi^{2k}(x).
\ee

The recursion for the odd state coefficients is
\bea
0 & = & (2j+1-2m)(j-m)d_{m-1} \non \\
& + & [E+(2m+1)(2j-2m-a-1)]d_m \non \\
& + & (2m+3)(m+1)d_{m+1},
\eea
with boundary conditions $ d_0=1 , \; d_{[j+ \half ]}=0 $.
The recursion for the odd energy polynomials $ q_{m} $ is
\bea
q_{m+1} & = & [E+(2m+1)(2j-2m-a-1)]q_m \nonumber \\
& - & (2j-2m+1)(j-m)(2m+1)mq_{m-1},
\eea
with initial conditions $ q_0=1 , \; q_1=E-2j $.
The relation between the polynomials and the coefficients is
\be
d_m= \frac{(-2)^m}{(2m+1)!}q_m(E)
\ee
The odd wave function is 
\be
\psi_n^{odd} = e^{-a(x)}\sum_{0}^{[j-\half]}\frac{q_k(E_n^{odd})}{(2k+1)!}\xi^{2k+1}(x)
\ee

The odd and even energy levels are interleaved, with the ground state always being even. 
The first polynomials and energy levels are given in table (\ref{termtable3}).
The lowest energy levels are very close because of tunneling between the two minima of the potential.
\begin{table}[h]
\begin{tabular}{|l||l|l||l|l|} \hline
$j$ & $Q^j(E)$ & $E_+$ & $q^j(E)$ & $E_-$ \\ \hline \hline
$0$ & $1$ & $-$ & $1$ & $-$ \\ \hline
$\half$ & $E$ & $0$ & $E-1$ & $1$ \\ \hline
$1$ & $E^2-2E-1$ & $\ba{l}
 -0.414214 \\ 2.41421
\ea$ & $E$ & $0$ \\ \hline
$\frac{3}{2}$ & $E^2-3$ & $\ba{c}
-1.73205 \\ 1.73205 
\ea$ & $E^2-2E-6$ & $\ba{c} -1.64575 \\ 3.64575 \ea$ \\ \hline
$2$ & $\ba{c} E^3-2E \\ -20E+24 \ea$ & $\ba{c} -4.17226 \\ 1.14399 \\ 5.02827 \ea$ & $E^2+2E-9$ & $\ba{c} -4.16228 \\ 2.16228 \ea$ \\ \hline 
$\frac{5}{2}$ & $E^3+4E^2-28E$ & $\ba{c} -7.65685 \\ 0 \\ 3.65685 \ea$ & $\ba{c} E^3+E^2 \\ -49E+15 \ea$ & $\ba{c} -7.65606 \\ 0.308667 \\  6.3474 \ea$ \\ \hline
$3$ & $\ba{c} E^4+4E^3 \\ -102E^2+12E \\ +585 \ea$ & $\ba{c} -12.150119 \\ -2.294220 \\ 2.715272 \\ 7.729066 \ea$ & $\ba{c} E^3+10E^2 \\ -36E-120 \ea$ & $\ba{c} -12.150071 \\ -2.246438 \\ 4.396509 \ea$ \\ \hline
$\frac{7}{2}$ & $\ba{c} E^4+16E^3 \\ -62E^2-528E \\ +945 \ea$ & $\ba{c} -17.645885 \\ -5.732662 \\ 1.623139 \\ 5.755408 \ea$ & $\ba{c} E^4+12E^3 \\ -144E^2-664E \\ +2100 \ea$ & $\ba{c} -17.645883 \\ -5.728016 \\ 2.286225 \\ 9.087674 \ea$ \\ \hline
\end{tabular}
\caption{\label{termtable3} The \qes sector energy levels of the hyperbolic cosine potential.}
\end{table}

%%%%%%%%%%%%%%%%%%%%%%%%%%%%%%%%%%%%%%%%%%%%%%%%%%%%%%%%%%%%
\section{Supersymmetric Potentials}
%%%%%%%%%%%%%%%%%%%%%%%%%%%%%%%%%%%%%%%%%%%%%%%%%%%%%%%%%%%%

\label{susy}
This section deals with supersymmetric (SUSY) potentials that can be represented by $ 2 \times 2 $ matrices or by Grassmann variables.
The wave functions are described either as two component spinors or as functions of two variables, one of which is Grassmann.
The super algebra discussed in this section is $ su(2/1) $, which is the simplest super algebra to give \qes potentials.
Other groups with a more complex structure can be chosen to give SUSY potentials that have more spinor components or more variables \cite{tanaka02}.
The $ su(2/1) $ algebra includes four even generators and four odd generators.
The even generators $ T^a,\; a=\pm ,0 $ are the generators of the $ su(2) $ algebra and $ J $ is the $ u(1) $ generator.
The odd generators are $ Q_\alpha $ and $ {\bar Q}_\alpha $ with $ \alpha=1,2 $.
The commutation and anti-commutation relations of these generators are given by
\bea
& & [ T^+,T^- ]=2T^0 ,\; [T^\pm,T^0]=\mp T^\pm ,\; [J,T^a]=0 \non \\
& & [ T^+,Q_1 ] = Q_2 ,\; [T^+,Q_2]=0 ,\; [T^-,Q_2]=Q_1 ,\; [T^-,Q_1]=0 \non \\
& & [ T^-,{\bar Q}_1 ] = {\bar Q}_2 ,\; [ T^-,{\bar Q}_2 ] = 0 ,\; [ T^+,{\bar Q}_2 ] = {\bar Q}_1 ,\; [ T^+,{\bar Q}_1 ] = 0 \non \\
& & [ Q_\alpha,J]=-\half Q_\alpha ,\; [{\bar Q}_\alpha,J]=+\half {\bar Q}_\alpha \non \\
& & \{ {\bar Q}_1,Q_2 \} =-T^+ ,\; \{ {\bar Q}_2,Q_1 \}=-T^- \non \\
& &  \half \{ {\bar Q}_1 ,Q_1-{\bar Q}_2,Q_2 \}=T^0 ,\; \half \{ {\bar Q}_1,Q_1+{\bar Q}_2,Q_2 \}=J
\eea
These generators can be realized by the operators
\be
\ba{ll}
\ba{lcl}
T^+ & = & 2j\xi - \xi ^2 \partial _\xi - \xi \theta \partial _\theta \\
T^0 & = & -j +\xi \partial _\xi + \half \theta \partial _\theta \\
T^- & = & \partial  _\xi \\
J & = & -j - \half \theta \partial _\theta
\ea & 
\ba{lcl}
Q & = & \left( \ba{c} Q_1 \\ Q_2 \ea \right) = \left( \ba{c} \partial _\theta \\ \xi \partial _\theta \ea \right) \\
{\bar Q} & = & \left( \ba{c} {\bar Q}_1 \\ {\bar Q}_2 \ea \right) = \left( \ba{c} \xi \theta \partial _\xi - 2j \theta \\ - \theta \partial _\xi \ea \right)
\ea
\ea.
\ee
$ \theta $ is a Grassmann variable realized by
\bea
\theta & = & \ket {\uparrow} = \left( \ba{c} 1 \\ 0 \ea \right) \non \\
\theta ^0 & = & \ket {\downarrow} = \left( \ba{c} 0 \\ 1 \ea \right),
\eea
or an operator realized by $ 2 \times 2 $ matrices
\bea
\theta & = & \sigma^+ = (\sigma_1 + i \sigma_2)/2 \non \\
\partial _\theta & = & \sigma^- = (\sigma_1 - i \sigma_2)/2,
\eea
where $ \sigma _i $ are the Pauli matrices
\be
\sigma_1 = \left( \ba{cc} 0 & 1 \\ 1 & 0 \ea \right), \;
\sigma_2 = \left( \ba{cc} 0 & -i \\ i & 0 \ea \right), \;
\sigma_3 = \left( \ba{cc} 1 & 0 \\ 0 & -1 \ea \right).
\ee
Through the same formalism as in the non SUSY case, the \schr equation is replaced by the equation
\be
H_G {\tilde \psi } = E {\tilde \psi }
\ee
where $ H_G $ is a a $ 2 \times 2 $ matrix which is a bilinear combination of all 8 generators denoted here by $ T^a $
\be
H_G = \sum_{a,b} c_{ab} T^a T^b +\sum_a c_aT^a,
\ee
and $ \tilde \psi $ is a 2-component wave function
\be
{\tilde \psi } = \left( \ba{c} {\tilde \psi }_u \\ {\tilde \psi }_d \ea \right) =  \left( \ba{c} \sum_{k=0}^{2j-1}u_k \xi^k \\ \sum_{k=0}^{2j}d_k \xi^k \ea \right).
\ee
This equation is invariant under the \qg transformation (in the SUSY potentials it is a matrix and not a scalar transformation). 

%%%%%%%%%%%%%%%%%%%%%%%%%%%%%%%%%%%%%%%%%%%%%%%%%%%%%%%%%%%%
\subsection{Example 8 - The Simplest Supersymmetric Potential}
%%%%%%%%%%%%%%%%%%%%%%%%%%%%%%%%%%%%%%%%%%%%%%%%%%%%%%%%%%%%

\label{susyexample}
The simplest known example of one-dimensional supersymmetric \qes potentials is given by the \qg Hamiltonian
\bea
H_G & = & - \{ T^0 , T^- \} - 2jT^- - {\bar Q}_2 Q_1 + \alpha T^0 \non \\
& - & i \beta (Q_2 T^- + {\bar Q}_1 + 2jQ_1) + \frac{i}{2}\alpha \beta Q_2 - \frac{i}{2}\beta Q_1.
\eea
As in example (\ref{simplest}), the change of variables is $ \xi(x)=x^2 $.
The potential is the $ 2 \times 2 $ matrix
\bea
V(x) & = & \frac{1}{8} (\alpha ^2 - \beta ^2)x^2 + \sigma_2 \left( -2j\beta -\frac{\beta}{4} + \frac{\alpha \beta x^2}{4} -\frac{\alpha}{4} \tan \frac{\beta x^2}{2} \right) \cos \frac{\beta x^2}{2} \non \\
& + &  \sigma_3 \left( -2j\beta -\frac{\beta}{4} + \frac{\alpha \beta x^2}{4} -\frac{\alpha}{4} \cot \frac{\beta x^2}{2} \right) \sin \frac{\beta x^2}{2}.
\eea
The new methods described in section (\ref{general}) are applied to this potential.
The \schr equation gives rise to wave functions with coefficients that satisfy the recursion relations
\bea
\label{susyrecursion}
0 & = & (2k+1)(k+1)d_{k+1} + \left[ E - \alpha \left( k - \frac{1}{4} \right) \right]d_k \non \\
& + & i \beta \left(2j + \half +k \right)u_k - \frac{i \alpha \beta}{2}u_{k-1} \non \\
0 & = & (2k+1)(k+1)u_{k+1} + \left[ E - \alpha \left( k + \frac{1}{4} \right) \right]u_k \non \\
& + & i \beta \left(2j - k \right)d_k,
\eea
with the boundary conditions $ d_{-1}=0 \; , \; u_{-1}=0\; , \; u_{2j}=0 \; , \;d_{2j+1}=0 $.
It is convenient to define the functions $ A_k, \; B_k, \; C_k, \; D_k, \; F_k, \; G_k, \; H_k $ as
\be
\label{susyabc}
\ba{ll}
\ba{rcl}
A_k & = & - \frac{i \alpha \beta}{2} \non \\
B_k & = & E - \alpha \left( k - \frac{1}{4} \right) \non \\
C_k & = & i \beta \left(2j + \half +k \right) \non \\
D_k & = & (2k+1)(k+1) \non \\
\ea
&
\ba{rcl}
F_k & = & i \beta \left(2j - k \right) \non \\
G_k & = & E - \alpha \left( k + \frac{1}{4} \right) \non \\
H_k & = & (2k+1)(k+1).
\ea
\ea
\ee
The relations (\ref{susyrecursion}) then can be written as
\bea
0 & = & A_ku_{k-1}+B_kd_k+C_ku_k+D_kd_{k+1} \non \\
0 & = & F_kd_k+G_ku_k+H_ku_{k+1}
\eea
The recursion relations for $ u_k $ and $ d_k $ can be separated to give a 4 term recursion for the upper wave function coefficients
\bea
0 & = & [-A_{k+1}F_{k+1}F_{k+2}]u_k\non \\
& + & [(B_{k+1}G_{k+1}-C_{k+1}F_{k+1})F_{k+2}]u_{k+1} \non \\
& + & [D_{k+1}F_{k+1}G_{k+2}+B_{k+1}F_{k+2}H_{k+1}]u_{k+2}\non \\
& + & [D_{k+1}F_{k+1}H_{k+2}]u_{k+3},
\eea
and a 4 term recursion for the lower wave function coefficients
\bea
0 & = & [A_{k+1}F_k(C_{k+2}G_{k+1}-A_{k+2}H_{k+1})]d_k \non \\
& + & [C_{k+2}F_{k+1}(C_{k+1}G_k-A_{k+1}H_k) \non \\
& + & B_{k+1}G_k(A_{k+2}H_{k+1}-C_{k+2}G_{k+1})]d_{k+1} \non \\
& + & [A_{k+1}B_{k+2}H_kH_{k+1}+G_k(-C_{k+2}D_{k+1}G_{k+1} \non \\
& + & H_{k+1}(A_{k+2}D_{k+1}-B_{k+2}C_{k+1}))]d_{k+2} \non \\
& + & [D_{k+2}H_{k+1}(A_{k+1}H_k-C_{k+1}G_k)]d_{k+3}.
\eea
The system of $ 4j+1 $ equations described in (\ref{susyrecursion}) can be written in matrix form.
Choosing the ordering of the variables to be
\be
{\vec C}_{4j+1} = \left( \ba{c} d_0 \\ u_0 \\ d_1 \\ u_0  \\ \vdots \\ d_{2j-1} \\ u_{2j-1} \\ d_{2j} \ea \right)
\ee
gives the $ (4j+1) \times (4j+1) $ matrix $ M_{4j+1} $ the form
\be
M_{4j+1} = \left( \ba{ccccccc}
B_0 & C_0 & D_0 & 0 & 0 & 0 & \ddots \\
F_0 & G_0 & 0 & H_0 & 0 & \ddots & 0 \\
0 & A_1 & B_1 & C_1 & \ddots & 0 & 0 \\
0 & 0 & F_1 & \ddots & 0 & H_{2j-2} & 0\\
0 & 0 & \ddots & A_{2j-1} & B_{2j-1} & C_{2j-1} & D_{2j-1} \\
0 & \ddots & 0 & 0 & F_{2j-1} & G_{2j-1} & 0 \\
\ddots & 0 & 0 & 0 & 0 & A_{2j} & B_{2j}
\ea \right),
\ee
where $ A, \; B, \; C, \; D, \; F, \; G, \; H $ are as defined in (\ref{susyabc}).
The \schr equation can be written as
\be
M_{4j+1} {\vec C} = 0
\ee 
The determinant of the coefficient matrix $ P_{4j+1}(E) = \det M_{4j+1} $ is the secular polynomial of the system and the zeros of this polynomial are the energy levels of the potential.
A recursion relation connecting the secular polynomials $ P_k(E) $ can be found using the technique used in section (\ref{secularp}).
This recursion is the four term relation
\bea
0 & = & [A_{k-1}D_{k-2} - B_{k-1}C_{k-2}] P_{2k+1} \non \\
& + & [-A_{k-1}B_kD_{k-2}G_{k-1}+C_{k-2}(A_kD_{k-1}F_{k-1} \non \\
& + & B_k(-C_{k-1}F_{k-1}+B_{k-1}G_{k-1}))] P_{2k-1} \non \\
& + & [A_{k-1}(-A_kD_{k-2}D_{k-1}G_{k-2}+B_k(B_{k-1}C_{k-2}H_{k-2} \non \\
& + & D_{k-2}(C_{k-1}G_{k-2}-A_{k-1}H_{k-2})))F_{k-1}] P_{2k-3} \\
& + & [A_{k-2}A_{k-1}D_{k-2}(B_kC_{k-1}-A_kD_{k-1})F_{k-1}H_{k-3}] P_{2k-5} \non
\eea
with 3 initial value conditions.

%%%%%%%%%%%%%%%%%%%%%%%%%%%%%%%%%%%%%%%%%%%%%%%%%%%%%%%%%%%%
\section{Summary of Results}
%%%%%%%%%%%%%%%%%%%%%%%%%%%%%%%%%%%%%%%%%%%%%%%%%%%%%%%%%%%%

In this chapter the exact solutions of some \qes potentials and some exactly solvable potentials are analyzed.
Solving the most general \qes potential in one dimension gives a set of recursion relations for the wave function coefficients.
The wave function coefficients are found to be proportional to the minors of the coefficient matrix as a result of applying Cramer's rule for the calculation of solutions of anharmonic systems of equations.
This result explains the connection between the boundary conditions and the quantization condition, which is the vanishing of the secular polynomial.
The recursion relation of the secular polynomials in the most general case is found to have more terms than the recursion relation of the wave function coefficients.
The different parities of \qes potentials are compared and the transition from even to odd parity potentials defined.
For all cases with $ \epsilon_k = 0 $ it is proved that the generating function of the secular polynomials solves the \schr equation with initial conditions.
The one-dimensional exactly solvable potentials are analyzed using the \qes formalism and the difference between exact solvability and quasiexact solvability is discussed.
For supersymmetric potentials the recursion relations for the upper and lower coefficients are separated, and a recursion relation is found for the secular polynomials.

%%%%%%%%%%%%%%%%%%%%%%%%%%%%%%%%%%%%%%%%%%%%%%%%%%%%%%%%%%%%
\chapter{Approximations}
%%%%%%%%%%%%%%%%%%%%%%%%%%%%%%%%%%%%%%%%%%%%%%%%%%%%%%%%%%%%

\label{approximations}
Typically, the \qes potentials have a free parameter $ j $ which determines the size of the \qes block.
For $ j \gg 1 $ the calculation of the energy levels is complicated.
A method of approximation is needed, which can simplify the calculation of the secular polynomials and energy levels for large $ j $.
In section (\ref{epsilonzero}) the generating function of the secular polynomials has been defined.
In this chapter the generating function is approximated using the \wkb approximation with initial value conditions (and not with boundary value conditions), and using the saddle point approximation.
A new approach to approximating the energy levels is described in this chapter.

%%%%%%%%%%%%%%%%%%%%%%%%%%%%%%%%%%%%%%%%%%%%%%%%%%%%%%%%%%%%
\section{\wkb Approximation of the Generating Function}
%%%%%%%%%%%%%%%%%%%%%%%%%%%%%%%%%%%%%%%%%%%%%%%%%%%%%%%%%%%%

\begin{figure}[p]
\epsfig{file=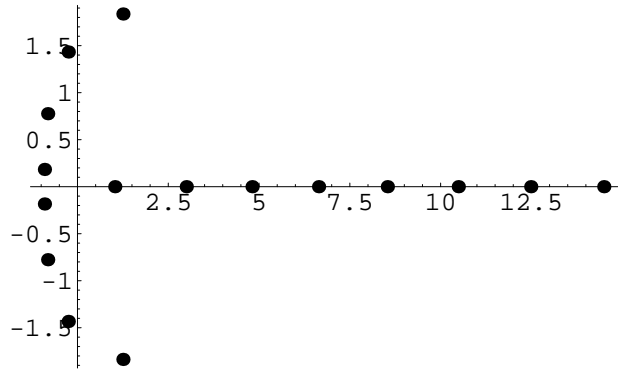}
\caption{\label{wkbgraph1} The results of differentiation of the harmonic oscillator first order \wkb approximation $ j= \frac{7}{2} $.}
\end{figure}

\begin{figure}[p]
\epsfig{file=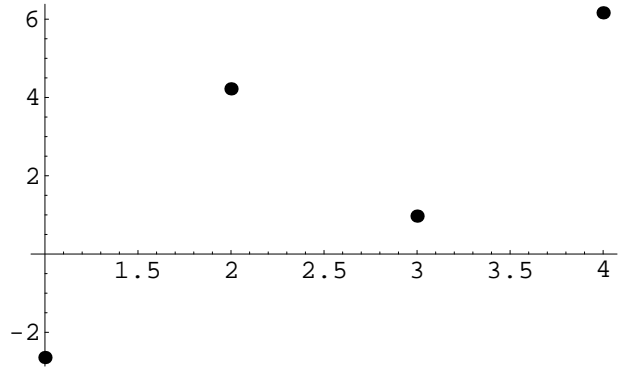}
\caption{\label{wkbgraph6} The normalized second highest energy of the sextic potential vs. the order of the \wkb approximation, $ j= \frac{7}{2} $.}
\end{figure}

In section (\ref{epsilonzero}) it is proved that for all cases with $ \eps_k=0 $ (\ref{abcde}) the initial value solution of the \schr equation is the generating function of the secular polynomials (\ref{generating4term})
\be
\Psi = e^{-a(x)} \sum_{k=0}^{\infty} \Delta_k P_k(E) \xi^k(x)
\ee
with initial conditions
\be
P_0(E) = 1 , \; P_1(E) = E
\ee
The secular polynomials can be calculated from $ \Psi $ by taking its $ k $th derivative in $ \xi $
\be
\label{derivatives}
P_k(E) = \left. \frac{1}{k!\Delta_k}\frac{d^k}{d\xi^k} \Psi(\xi(x)) \right| _{\xi=0}
\ee
The secular polynomials and energy levels are approximated in this chapter using the \wkb approximation with initial value conditions instead of boundary value conditions (Appendix \ref{wkbapp}).
The solutions are determined by two free parameters
\bea
\label{psiwkb1}
\psi^{\wkb} & = & \frac{A(E)}{\sqrt[4]{E-V(x)}} \; \exp \left[ i \sqrt{2} \int_0^x \sqrt {E-V(y)} \; dy\right] \non \\
& + & \frac{B(E)}{\sqrt[4]{E-V(x)}} \; \exp \left[ -i \sqrt{2} \int_0^x \sqrt {E-V(y)} \; dy \right].
\eea
Initial value conditions have the form $ \Psi (x_0) = a , \; \frac{d}{dx}\Psi (x_0) = b $. For the examples discussed in this chapter the generating function has a well defined parity and therefore the choice of 
\be
\left\{ \ba{l}
\psi(0,E)=1\\
\psi(-x,E)=\psi(x,E)
\ea  \right. \Rightarrow \left\{ \ba{l}
\psi(0,E)=1\\
\psi '(0,E)=0
\ea  \right.
\ee
as the initial value conditions is obvious.
The \wkb approximation of the generating function for even energy levels is
\be
\psi^{\wkb} = \sqrt[4]{\frac{E}{E-\frac{x^2}{2}}} \: \cos \left[\sqrt 2 \int_0^x \sqrt{E-\frac{y^2}{2}} \: dy \right].
\ee
The approximate generating function is now used instead of the exact function in equation (\ref{derivatives}) to give the approximate secular polynomials
\be
\label{wkbderivatives}
P_k(E) = \left. \frac{1}{k!\Delta_k}\frac{d^k}{d\xi^k(x)} \psi^{\wkb}(\xi(x)) \right| _{\xi=\xi(x=0)}.
\ee
The results of this approximation were not good enough so the computation was repeated with higher order \wkb approximations.
The results improved with the order of \wkb but the complexity of the calculation increased so that large $ j $ calculation was still out of reach.
The results of these approximations are presented in table (\ref{wkbtable}) and in figures (\ref{wkbgraph1}-\ref{wkbgraph6})
\begin{table}[h]
\begin{tabular}{|c||c|c|c|} \hline
Potential & Exact & First Order \wkb & Fourth Order \wkb \\ \hline \hline
$ \ba{c} \mbox{Harmonic} \\ \mbox{Oscillator} \\ j= \frac{7}{2} \ea $ & $ \ba{c} 14.5 \\ 12.5 \\ 10.5 \\8.5 \\ 6.5 \ea $ & $ \ba{c} 14.493 \\12.482 \\10.487 \\ 8.533 \\ 6.639 \ea $ & $ \ba{c} 14.5 \\ 12.5001 \\10.5009 \\8.5004 \\ 6.471 \ea $ \\ \hline
$ \ba{c} \mbox{Sextic} \\ \mbox{Potential} \\ j= \frac{7}{2} \ea $ & $ \ba{c} 27.804 \\ 17.777 \\ 9. 041 \\ 2.273 \\ -2.273 \ea $  &  $ \ba{c} 26.711 \pm i 4.86 \\ 15.134 \pm i 11.046 \\ 6.455 \pm i 12.927 \\ \pm i 12.866 \\ \pm i 5.684 \ea $ & $ \ba{c} 29.168 \\ 23.94 \pm i 7.187 \\ 16.012 \pm i 13.78 \\ 7.85 \pm i 16.39 \\ 5.398 \ea $ \\ \hline
\end{tabular}
\caption{\label{wkbtable} The exact and approximate energy levels of the harmonic oscillator and sextic potentials, in first and fourth order \wkb approximation.}
\end{table}
\begin{figure}[p]
\epsfig{file=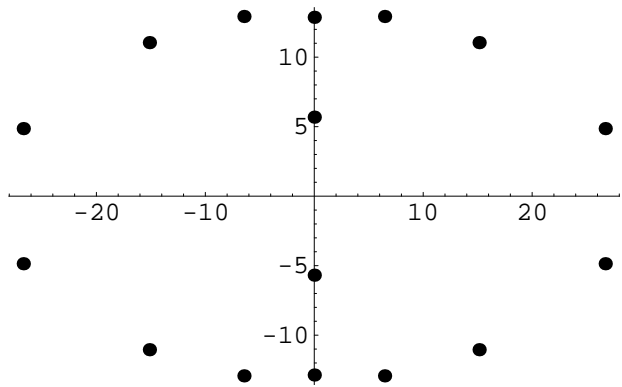}
\caption{The results of differentiation of the sextic potential first order \wkb approximation $ j= \frac{7}{2} $.}
\end{figure}
\begin{figure}[p]
\epsfig{file=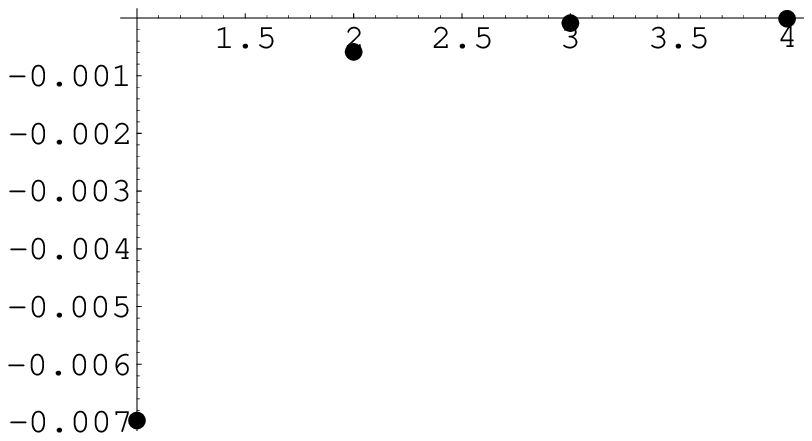}
\caption{The normalized highest energy of the harmonic oscillator vs. the order of the \wkb approximation, $ j= \frac{7}{2} $.}
\end{figure}
\begin{figure}[p]
\epsfig{file=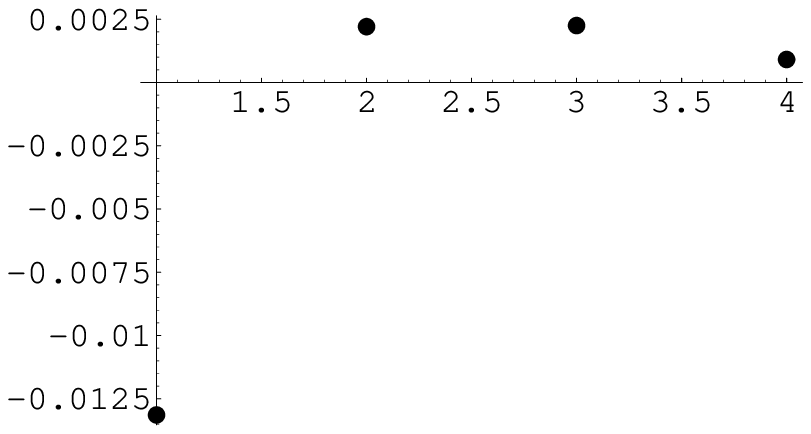}
\caption{The normalized third highest energy of the harmonic oscillator vs. the order of the \wkb approximation, $ j= \frac{7}{2} $.}
\end{figure}
\begin{figure}[p]
\epsfig{file=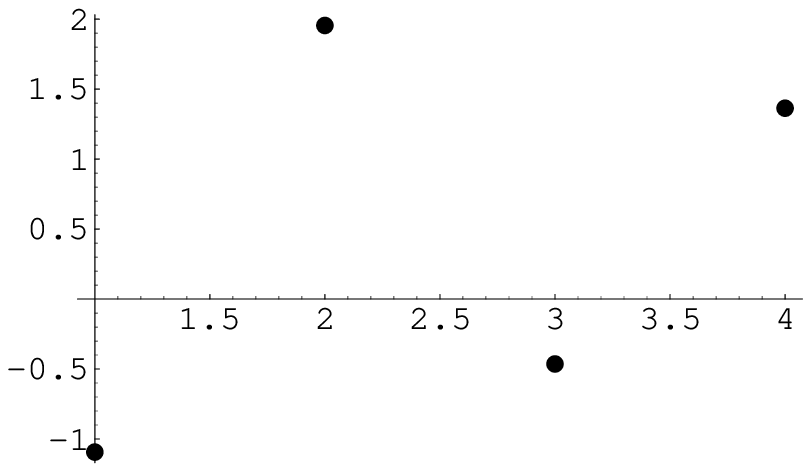}
\caption{The normalized highest energy of the sextic potential vs. the order of the \wkb approximation, $ j= \frac{7}{2} $.}
\end{figure}

%%%%%%%%%%%%%%%%%%%%%%%%%%%%%%%%%%%%%%%%%%%%%%%%%%%%%%%%%%%%
\section{Saddle Point Approximation of The Secular Polynomials}
%%%%%%%%%%%%%%%%%%%%%%%%%%%%%%%%%%%%%%%%%%%%%%%%%%%%%%%%%%%%

The calculation of energy levels through equation (\ref{wkbderivatives}) requires very long computations.
The fact that the interesting limit is $ j \gg 1 $ can be used to apply the saddle point approximation on the Cauchy integral
\be
\label{cauchy}
P_n(E)=\frac{1}{\Delta_n} \oint _{c_0} \frac{d\xi(x)}{2 \pi i} \frac{e^{a(x)}}{\xi^{n+1}(x)} \Psi^{\wkb} (\xi(x)),
\ee
which is equivalent to equation (\ref{wkbderivatives}).
For the saddle point approximation of the Cauchy integrals the variables have to be changed into new variables that are scaled with $ j $, so that the exponential phase is proportional to $ j $.
This calculation has many details that are difficult to generalize, so two examples are given below.

%%%%%%%%%%%%%%%%%%%%%%%%%%%%%%%%%%%%%%%%%%%%%%%%%%%%%%%%%%%%
\subsection{Secular Polynomials of the Harmonic Oscillator}
%%%%%%%%%%%%%%%%%%%%%%%%%%%%%%%%%%%%%%%%%%%%%%%%%%%%%%%%%%%%

\begin{figure}[p]
\epsfig{file=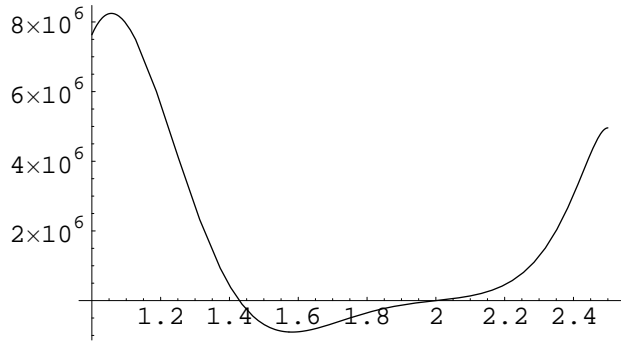}
\caption{\label{hosaddler2} Real part of saddle point approximated harmonic oscillator secular polynomial.}
\end{figure}

\begin{figure}[p]
\epsfig{file=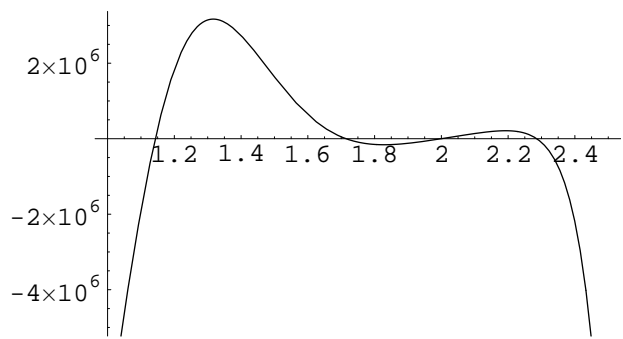}
\caption{\label{hosaddlei2} Imaginary part of saddle point approximated harmonic oscillator secular polynomial.}
\end{figure}

\label{hospa}
The harmonic oscillator is chosen to demonstrate this approximation because all the energy levels are known exactly.
In section (\ref{hoe}) it is shown that for the potential
\be
V(x) = \half x^2
\ee
the initial value solution of the \schr equation is
\be
\Psi = e^{-\frac{x^2}{2}} \sum_{k=0}^{\infty} \frac{(-2)^k}{(2k)!} P_k(E)x^{2k}
\ee
where
\be
P_k(E) = \prod_{i=0}^{k-1} \left( E-2i-\half \right)
\ee
are the secular polynomials.
The zeros of the polynomial $ P_{2j+1}(E) $ give the first $ 2j+1 $ even energies of the system
\be
E_n^{even} = 2n + \half , \; n = 0,1,2,...
\ee
The approximate \wkb initial value solution (\ref{psiwkb1}) for the harmonic oscillator is
\be
\psi^{\wkb} = \sqrt[4]{\frac{E}{E-\frac{x^2}{2}}} \: \cos \left[\sqrt 2 \int_0^x \sqrt{E-\frac{y^2}{2}} \: dy \right].
\ee
Inserting this definition into the Cauchy integral (\ref{cauchy}) gives
\be
P_k(E) \propto \oint _{c_0} \frac{dx}{2 \pi i} \frac{e^{\frac{x^2}{2}}}{x^{2k+1}} \sqrt[4]{\frac{2E}{2E-x^2}} \; \cos \left[ \int_0^x \sqrt{2E-y^2} \: dy \right]
\ee
We are interested in the $ 2j+1 $ polynomial.
For the proceeding calculations it is more convenient to work with the exponential form of the cosine
\be
\label{pwkbcuachy}
P_{2j+1}(E) \propto \sum _{\alpha=\pm 1} \oint _{c_0} \frac{dx}{2 \pi i} \frac{e^{\frac{x^2}{2}}}{x^{4j+3}} \sqrt[4]{\frac{2E}{2E-x^2}} \; e^{ \alpha i \int_0^x \sqrt{2E-y^2} \: dy }.
\ee
In order to do a saddle point approximation of this integral we need a single large parameter ($ j $) that multiplies the entire phase.
To this end we will change the variables to scaled variables.
The choice of the scaling of the variables is such that all the terms in the exponential phase will be linearly dependent on $ j $.
The scaled variables are
\bea
x & = & \sqrt{j} \; s \non \\
y & = & \sqrt{j} \; t \non \\
E & = & j \; \epsilon
\eea
Substituting the scaled variables in (\ref{pwkbcuachy}) results in
\be
\label{pwkb1}
P_{2j+1}(\epsilon) \propto \sum_{\alpha = \pm 1} \oint_{c_0} \frac{ds}{2\pi is^3}\sqrt[4]{\frac{\epsilon}{\epsilon - \frac{s^2}{2}}} \; e^{jS_\alpha(s)}
\ee
where the phase $ S_\alpha (s) $ is
\be
S_\alpha(s) = \alpha i\sqrt 2 \int_0^s \sqrt{\epsilon - \frac{t^2}{2}} \; dt + \frac{s^2}{2} - 4\ln s.
\ee
This integral will be the starting point of the saddle point approximation.
The dominant contribution to the integral (\ref{pwkb1}) is from constant phase paths that pass through the saddle points \cite{benderorzag}.
The phase $ S_\alpha(s) $ is expanded to a second order Taylor series around the saddle points $ s_\beta $
\be
S_\alpha(s) = S_\alpha(s_\beta + z) = S_\alpha(s_\beta) + S_\alpha'(s_\beta)z + \half S_\alpha''(s_\beta)z^2.
\ee
The saddle points satisfy
\be
S_\alpha'(s) = \alpha i\sqrt 2 \sqrt{\epsilon - \frac{s^2}{2}} + s - \frac{4}{s} = 0,
\ee
and are equal to
\be
\label{sbeta}
s_\beta(\epsilon) = \beta \sqrt{\frac{8}{4-\epsilon}} \; \; , \; \; \beta = \pm 1.  
\ee
The values of the phase and its derivatives at the saddle points are
\bea
S_\alpha(s_\beta) & = & \alpha i\sqrt 2 \int_0^{\beta \sqrt{\frac{8}{4-\epsilon}}} \sqrt{\epsilon - \frac{t^2}{2}} \; dt + \frac{8}{4-\epsilon} - 2 \ln \left( \frac{8}{4-\epsilon} \right) \non \\
S_\alpha'(s_\beta) & = & 0 \non \\
S_\alpha''(s_\beta) & = & - \frac{(4-\epsilon)^2}{2(\epsilon-2) }.
\eea
We are interested in paths of steepest descents so we want $  \half S_\alpha''(s_\beta)z^2 $ to be real and negative.
Since $  S_\alpha''(s_\beta) $ is negative, $ z $ is chosen to be real.
\begin{figure}[h]
\epsfig{file=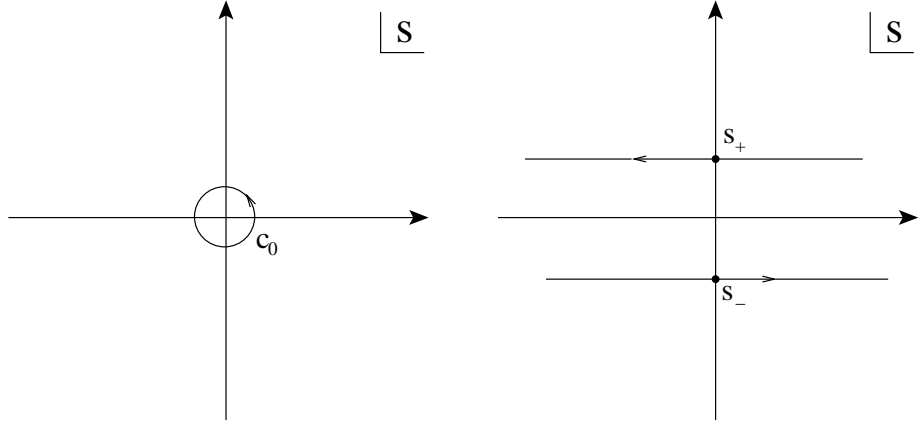}
\caption{\label{splane} Two paths of integration in the s-plane for the saddle point approximation of the harmonic oscillator secular polynomials.}
\end{figure}
The dominant contribution to the integral (\ref{pwkb1}) is therefore the steepest descent integral
\be
P_{2j+1}(\epsilon) \propto \sum_{\alpha,\beta = \pm 1} \sqrt[4]{\frac{\epsilon}{\epsilon - \frac{s_\beta^2}{2}}} \; \frac{e^{jS_\alpha(s_\beta)}} {2\pi i s_\beta^3} \int_0^\infty dz \; e^{-j \frac{(4-\epsilon)^2}{2(\epsilon-2) } z^2 }.
\ee
After integrating and substituting $ s_\beta $ from (\ref{sbeta}) the approximate polynomial is
\be
P_{2j+1}(\epsilon) \propto \sum_{\alpha,\beta = \pm 1} (4-\epsilon)^{2j} \; e^{\frac{4j}{4-\epsilon}} \; \frac{\sqrt[4]{\epsilon(4-\epsilon)}} {\sqrt{\alpha \beta j}} \; e^{ \alpha ij \int_0^{\beta \sqrt{\frac{8}{4-\epsilon}}} \sqrt{2\epsilon - t^2} dt }.
\ee
Using the parity of the potential to sum over $\alpha$ and $\beta$ the final result of the saddle point approximation of the secular polynomial is
\be
\label{finalpolynomial}
P_{2j+1}(\epsilon) \propto (4-\epsilon)^{2j}e^{\frac{4j}{4-\epsilon}} \sqrt[4]{\epsilon(4-\eps)} \cos \left[j \int_0^{\sqrt{\frac{8}{4-\epsilon}}} \sqrt{2\epsilon - t^2} \; dt \right].
\ee
The real and imaginary parts of $ P_{2j+1}(\epsilon) $ are plotted in figures (\ref{hosaddler2}) and (\ref{hosaddlei2}) respectively.
The approximate energy levels are given by the zeros of the approximate secular polynomial.
$ P_{2j+1}(\eps) $ vanishes for $ \eps = 0 $, which corresponds to the lowest exact energy $ E_0^{even}=\half $.
The limit of (\ref{finalpolynomial}) for $ \eps \ra 4 $ diverges.
The cosine has a zero for $ \eps = 2 $ and other zeros, all of which solve the equation
\be
\label{saddlefinal}
j \int_0^{\sqrt{\frac{8}{4-\epsilon}}} \sqrt{2\epsilon - t^2}  \; dt = \left(k+\half \right)\pi.
\ee
Expanding the integral gives
\be
\label{saddlephase}
\int_0^{\sqrt{\frac{8}{4-\epsilon}}} \sqrt{2\epsilon - t^2} dt = \pi + \frac{\pi}{2} (\eps - 2) - \frac{i}{6} (\eps-2)^3 - \frac{i}{12} (\eps-2)^4 + \cdot \cdot \cdot
\ee
The real part of this number is the integral over the allowed region $ \int_0^{\sqrt{2 \epsilon}} \sqrt{2\epsilon - t^2} \; dt $.
Solving equation (\ref{saddlefinal}) in the highest order in $ \frac{1}{j} $ results in
\be
k = j
\ee

This result is valid only to first order (because the previous approximation was in leading order only), which means that the final result of the approximation is
\be
E_{k}^{even} = 2k
\ee
which for large $ k $ is an excellent approximation.
This result can be improved by computing the approximations to higher orders, both in the \wkb approximation and in the saddle point approximation.

%%%%%%%%%%%%%%%%%%%%%%%%%%%%%%%%%%%%%%%%%%%%%%%%%%%%%%%%%%%%
\subsection{The Bender-Dunne Polynomials}
%%%%%%%%%%%%%%%%%%%%%%%%%%%%%%%%%%%%%%%%%%%%%%%%%%%%%%%%%%%%

\begin{figure}[p]
\epsfig{file=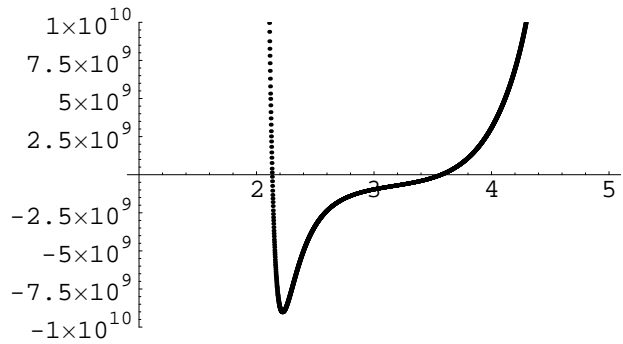}
\caption{\label{bdr} Real part of saddle point approximated Bender-Dunne polynomial. $ j=\frac{7}{2} $.}
\end{figure}

\begin{figure}[p]
\epsfig{file=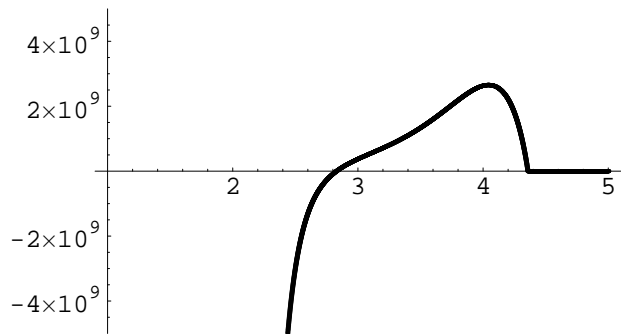}
\caption{\label{bdi} Imaginary part of saddle point approximated Bender-Dunne polynomial. $ j=\frac{7}{2} $.}
\end{figure}

The calculation of section (\ref{hospa}) are repeated for the sextic potential (\ref{simplestpotential}), which is the simplest most popular example of \qes potentials
\be
V(x) = \half x^6 - \half(8j+3)x^2
\ee
The generating function of the secular polynomials of this potential is
\be
\Psi = e^{-\frac{x^2}{2}} \sum_{k=0}^{\infty} \frac{(-2)^k}{(2k)!} P_k(E)x^{2k}
\ee
where the secular polynomials $ P_k(E) $ are given by the recursion relation
\be
P_{k+1}(E)=EP_k(E)-2k(2k-1)(2j+1-k)P_{k-1}(E)
\ee
with the initial conditions $ P_0=1 , \; P_{1}=E $. 
The first energy levels and secular polynomials are listed in table (\ref{x^6}).
The \wkb approximated generating function is 
\be
\label{wkb}
\psi^{\wkb} = \sqrt[4]{ \frac{E}{E-V(x)}} \cos \left[ \sqrt{2} \int _0 ^x \sqrt {E-V(y)} \; dy \right]
\ee
Incorporating this into (\ref{cauchy}), taking $ k=2j+1 $ and writing the cosine as a sum of integrals gives
\be
\label{pje1}
P_{2j+1} (E) \propto \sum_{\alpha = \pm 1} \oint _{c_0} \frac{dx}{2 \pi i} \frac{e^{\frac{x^4}{4}}}{x^{4j+3}} \sqrt[4]{ \frac{E}{E-V(x)}} e^{ \alpha i \sqrt{2} \int _0 ^x \sqrt {E-V(y)} dy }.
\ee
The contour integrals can be approximated using the saddle points method for large $ j $.
The saddle point approximation requires us to define scaled variables so that the exponential phase is linearly dependent on $ j $
\bea
x & = & j^{\frac{1}{4}} s \non \\
y & = & j^{\frac{1}{4}} t \non \\
E & = & j^{\frac{3}{2}} \epsilon \non \\
V(x) & = & \half x^6 - \half (8j+3) x^2 =  j^{\frac{3}{2}} [\half s ^6 -4 s ^2].
 =  j^{\frac{3}{2}} v(s)
\eea
Substituting these variables in $ P_{2j+1} (E) $ leads to
\be
P_{2j+1} (\epsilon) \propto \sum_{\alpha = \pm 1} \oint _{c_0} \frac{ds}{2 \pi i s ^3} \sqrt[4]{\frac{\epsilon}{\epsilon-v(s)} } \; e^{jS_{\alpha}}
\ee
where the phase $ S_\alpha $ is
\be
S_\alpha = \alpha i \sqrt{2} \int _0 ^s \sqrt {\epsilon - v(t)} \; dt + \frac{s^4}{4} -4 \ln s.
\ee
This is the starting point of the saddle point approximation.
To find the integration paths the phase $ S_\alpha(s) $ is now expanded in a Taylor series around
 the saddle points $ s_\beta $
\be
S_\alpha (s) = S_\alpha (s_\beta (\epsilon) + z) =  S_\alpha (s_\beta (\epsilon)) + \half S''_\alpha (s_\beta (\epsilon))z^2.
\ee
The saddle points are the points where the first derivative vanishes
\be
\frac{d}{ds}S_\alpha = \alpha i \sqrt{2} \sqrt {\epsilon - v(s)} + s^3 - \frac {4}{s} =0.
\ee
The saddle points of the integral (\ref{pje1}) are
\be
\label{sp}
s_\beta (\epsilon) = \beta i \; \frac{2\sqrt{2}}{\sqrt{\epsilon}}, \; \beta = \pm 1.
\ee
The values of $ S_\alpha $ and its derivatives at the saddle points are
\bea
S_\alpha(s_\beta(\epsilon)) & = & \alpha i \sqrt{2} \int _0 ^{s_\beta(\epsilon)} \sqrt {\epsilon - v(t)} \; dt +\frac{16}{\epsilon ^2}  - \ln (\frac {64}{\epsilon ^2}) \\
\frac{d}{ds}S_\alpha(s_\beta(\epsilon)) & = & 0 \\
\frac{d^2}{ds^2 }S_\alpha(s_\beta(\epsilon)) & = & \frac{- \epsilon^3}{2 \left(\epsilon ^2 - 16 \right)}.
\eea
The range of epsilon for the \qes sector is
\be
- \frac{16}{3}\sqrt{\frac{2}{3}} < \epsilon < \frac{16}{3}\sqrt{\frac{2}{3}},
\ee
The second derivative must be real and negative for paths of steepest descents. 
Since the second derivative changes sign at $ \eps = \pm 4 $ and the interesting energies are the highest, the range is limited to
\be
4 < \epsilon < \frac{16}{3} \sqrt{\frac{2}{3}} \simeq 4.35
\ee
though the final result should be the same for the entire range of $ \eps $.
For the chosen range the path parameter $ z $ has to be real and the integration paths are plotted in (\ref{splane}).
The main contribution to the value of $ P_{2j+1}(\epsilon) $ is therefore from the integrals
\be
P_{2j+1}(\epsilon) \propto \sum_{\alpha,\beta=\pm 1} \frac{e^{jS_\alpha(s_\beta)}} {2 \pi i s_\beta ^3} \left( \frac{\epsilon} {\epsilon-v(s_\beta)} \right) ^{\frac{1}{4}}  \int _0^\infty dz \; e^{-j \frac{\epsilon^3}{4 \left(\epsilon - 16 \right)}z^2 }.
\ee
After the integration, $ s_\beta $ of (\ref{sp}) is substituted and the parity of the potential is used to get the approximate value of $ P_{2j+1}(\epsilon) $
\be
P_{2j+1}(\epsilon) \propto \eps ^{2j+1} e^{j \frac {16}{\epsilon^{2}}} \; \cos \left[ j \sqrt{2} \int _0 ^{i \frac{2\sqrt{2}}{\sqrt{\epsilon}}} \sqrt {\epsilon - v(t)} \;  dt \right].
\ee
The approximate polynomial can be evaluated numerically as shown in figures (\ref{bdr}-\ref{bdi}).
The zeros of this function are the zeroes of the cosine, which can be calculated order by order in $ j $.
The leading order calculation results in
\be
E_* =  \frac{16}{3} \sqrt{\frac{2}{3}} \; j^{\frac{3}{2}},
\ee
which is the approximate highest energy of the \qes sector obtained in \cite{benderdunnemoshe}.
The numerical results are summed in table (\ref{bdspatable}).
This is a first order approximation of the highest \qes energy that is obtained through a general calculation independent of the energy reflection symmetry.
\begin{table}[h]
\begin{tabular}{|l||l|l|} \hline
$ j $ & Exact & Saddle Point Approximation \\ \hline \hline
$ \frac{15}{2} $ & $ 88.404 $ & $ 89.443 $ \\ \hline
$ 8 $ & $ 97.461 $ & $ 98.534 $ \\ \hline
$ \frac{17}{2} $ & $ 106.809 $ & $ 107.915 $ \\ \hline
$ 9 $ & $ 116.438 $ & $ 117.576 $ \\ \hline
$ \frac{19}{2} $ & $ 126.34 $ & $ 127.508 $ \\ \hline
$ 10 $ & $ 136.507 $ & $ 137.706 $ \\ \hline
\end{tabular}
\caption{\label{bdspatable} The exact and saddle point approximate energy levels of the harmonic oscillator.}
\end{table}

%%%%%%%%%%%%%%%%%%%%%%%%%%%%%%%%%%%%%%%%%%%%%%%%%%%%%%%%%%%%
\section{Summary of Results}
%%%%%%%%%%%%%%%%%%%%%%%%%%%%%%%%%%%%%%%%%%%%%%%%%%%%%%%%%%%%

The \wkb approximation of the generating function is a new approach to the approximation of energy levels.
Combined with the saddle point approximation the complexity of the energy calculation is small.
This calculation is a first order approximation of the energy levels of the \qes, potentials to leading order in $ \frac{1}{j} $.
For the sextic potential the calculation approximates the highest \qes sector energy level as a function of $ j $ without any dependence on the energy reflection symmetry.

%%%%%%%%%%%%%%%%%%%%%%%%%%%%%%%%%%%%%%%%%%%%%%%%%%%%%%%%%%%%
\chapter{Summary and Conclusions}
%%%%%%%%%%%%%%%%%%%%%%%%%%%%%%%%%%%%%%%%%%%%%%%%%%%%%%%%%%%%

\label{summary}
\Qe solvable potentials have several unexplored properties.
This work concentrates on the connection between initial value and boundary value solutions of the one-dimensional \schr equation, which is then used as a basis for approximation.

\Qg Hamiltonians in one dimension can be constructed from any bilinear combination of the generators of the $ sl(2) $ Lie group. 
The basis of the solutions is constructed from the $j$th representation of the group and an arbitrary set that spans the orthogonal space. 
As a result the \qg Hamiltonian has a block diagonal form with at least one finite block which has the same dimension as the representation ($ 2j+1 $). 
The wave function that solves the \schr equation has an exponential factor and a pre-exponential function which is a polynomial of degree $ 2j+1 $ in the group variable $ \xi $. 
The \schr equation may be replaced by a recursion relation for the wave function coefficients.

In chapter (\ref{exact}) exact solutions of some \qes potentials and some exactly solvable potentials are discussed. 
In section (\ref{general}) a new solution of the most general one-dimensional \qg Hamiltonian is constructed.
This solution includes the transformation between the Hamiltonian and the \qg Hamiltonian $ H_G $ by calculating the change of variables, the exponential behavior of the wave functions and the potential.
The recursion relations for the wave function coefficients is written in matrix form. 
The quantization condition for this system is given by the vanishing of the secular polynomial, which is the determinant of the coefficient matrix.
The secular polynomials are found to be related to each other, through a general recursion relation, with initial conditions.
This simplifies the computation of the energy levels for any one-dimensional \qes potential.
The coefficients of the pre-exponential part of the wave function are found to be proportional to the minors of the coefficient matrix, which explains the connection between the boundary value conditions and the quantization condition.
The connection between the even and odd parity potentials of the same family (i.e. with the same $ \xi(x) $) is shown.
In section (\ref{epsilonzero}) it is proved that for all \qes potentials with $ \eps_k = 0 $, the generating function of the secular polynomials is the initial value solution of the \schr equation, thus expanding the results of existing works.
In section (\ref{exactlysolvable}) the one-dimensional exactly solvable potentials are analyzed using \qes formalism and the difference between exact solvability and \qe solvability is discussed.
In section (\ref{susy}) new recursion relations for the secular polynomials of the SUSY potential are derived.  
The separate recursion relations for coefficients of the upper and lower parts of the spinor wave function are found.

In chapter (\ref{approximations}) the exact energy levels of \qes potentials are approximated using a new technique in order to simplify the calculation for large $ j $.
The generating function of the secular polynomials is approximated using the \wkb approximation with initial value conditions (in contrast to the more common boundary value \wkb solution).
The approximate energy levels are derived by replacing the exact generating function by the approximate function.
Since the complexity of this calculation is considerable, a second approximation is introduced.
The derivatives of the generating functions are replaced by Cauchy integrals, which are then calculated using the saddle point approximation.
This method is demonstrated for two specific potentials - the harmonic oscillator and the sextic potential.
The zeros of the approximate secular polynomials (which are the approximate energy levels) are calculated order by order in the group parameter $ \frac{1}{j} $.
For the harmonic oscillator, for an energy of order $ j $ the error is of order 1.
The results are valid for energies that are not in the top part of the calculated spectrum, because the approximate secular polynomial has singular points for high energy levels. 
For the sextic potential only the highest energy of the \qes part of the spectrum can be evaluated.
For the highest energy of order $ j^{3/2} $ the error is of order $ j^{1/2} $.
This result was obtained before by numerical calculations and proved analytically based on the energy reflection symmetry.
Using our approximation this result can be duplicated for any potential of interest without this symmetry.

This work can be continued by generalizing the calculations to include the higher dimensions and other groups, including SUSY groups.
The approximation can be applied to more potentials and improved by introducing the saddle point approximation to higher order, with higher order \wkb approximation.

%%%%%%%%%%%%%%%%%%%%%%%%%%%%%%%%%%%%%%%%%%%%%%%%%%%%%%%%%%%%
\appendix
%%%%%%%%%%%%%%%%%%%%%%%%%%%%%%%%%%%%%%%%%%%%%%%%%%%%%%%%%%%%

%%%%%%%%%%%%%%%%%%%%%%%%%%%%%%%%%%%%%%%%%%%%%%%%%%%%%%%%%%%%
\chapter{Oscillation Theory}
%%%%%%%%%%%%%%%%%%%%%%%%%%%%%%%%%%%%%%%%%%%%%%%%%%%%%%%%%%%%

\label{oscillation}
In one-dimensional potentials the energy levels are not degenerate and therefore can be ordered by magnitude.
The wave functions can be numbered according to the number of zeros (or nodes).
The two numbering methods coincide.

In the solution of the one-dimensional eigenvalue problem
\be
H(x) \psi(x) = - \psi''(x) + V(x) \psi(x) =  E \psi(x)
\ee
there is no degeneracy of the energy levels, since if two different solutions $ \psi_1 $ and $ \psi_2 $
\bea
\label{beginapp1}
- \psi''_1 + V \psi_1 & = & E_1 \psi_1 \; \; \; \\
- \psi''_2 + V \psi_2 & = & E_2 \psi_2 \; \; \;
\eea
are assumed to have the same energy
\be
E_1=E_2,
\ee
then the difference $ \psi''_1 \psi_2 - \psi_1 \psi''_2 $ will be equal to zero
\bea
- \psi''_1 \psi_2 + V \psi_1 \psi_2 & = & E \psi_1 \psi_2 \\
- \psi''_2 \psi_1 + V \psi_2 \psi_1 & = & E \psi_2 \psi_1 \\
- \left( \psi''_1 \psi_2 - \psi_1 \psi''_2 \right) & = & - \frac{d}{dx} \left( \psi'_1 \psi_2 - \psi'_2 \psi_1 \right) =0,
\label{endapp1}
\eea
which means that the Wronskian $ W(x) $
\be
W(x) = \psi'_1 \psi_2 - \psi'_2 \psi_1 = const
\ee
is constant.
The Wronskian must vanish for $ x \rightarrow \pm \infty $ so it is constantly zero
\be
W(x) \stackrel {x \rightarrow \pm \infty} {\longrightarrow} = 0 \: \Rightarrow \: W(x)=0 \: \Rightarrow \: E_1 \neq E_2.
\ee
This means that the two wave functions must be identical or the energies not equal.

Repeating the calculation (\ref{beginapp1}-\ref{endapp1}), with $  E_2 > E_1 $ results in
\be
\label{wronapp1}
- \frac{d}{dx} \left( \psi'_1 \psi_2 - \psi'_2 \psi_1 \right) = (E_1-E_2) \psi_1 \psi_2.
\ee
If $ \psi_1 $ has zeros in $ x=a,b $ and is positive in the section $ (a,b) $ then integrating (\ref{wronapp1}) from $ a $ to $ b $ leads to
\be
\label{intapp1}
\left. - \left( \psi'_1 \psi_2 - \psi'_2 \psi_1 \right) \right|_a^b = (E_1-E_2) \int_a^b \psi_1 \psi_2 \; dx.
\ee
Since $ \psi_1(a)=0 , \; \psi_1(b)=0 , \; \psi_1(a<x<b)>0 $, the derivatives of $ \psi_1 $ at the end points are known to be $ \psi'_1(a)>0 , \; \psi'_1(b)<0 $.
If we assume $ \psi_2(a)>0 $ then the left hand side of (\ref{intapp1}) is positive
\be
{\cal L} = \psi'_1(a) \psi_2(a) - \psi'_1(b) \psi_2(b) > 0,
\ee
and if we also assume that $ \psi_2(a \le  x \le b)>0 $ then the right hand side is negative
\be
{\cal R} = (E_1-E_2) \int_a^b \psi_1 \psi_2 \; dx < 0,
\ee
which is a contradiction. The assumption that $ \psi_2(a \le  x \le b)>0 $ is proved to be wrong and therefore $ \psi_2 $ has a zero in the section  $ (a,b) $.
Since at the boundary both wave functions vanish, $ \psi_2 $ must have one zero more than $ \psi_2 $.
Since the ground state has no zeros, the number of nodes of the n-th eigenfunction (corresponding to the n-th eigenvalue) is n.
A more thorough discussion on oscillation theory can be found for example in \cite{landaulifshitz,couranthilbert}.

%%%%%%%%%%%%%%%%%%%%%%%%%%%%%%%%%%%%%%%%%%%%%%%%%%%%%%%%%%%%
\chapter{Initial Value and Boundary Value}
%%%%%%%%%%%%%%%%%%%%%%%%%%%%%%%%%%%%%%%%%%%%%%%%%%%%%%%%%%%%

\label{initialboundary}
A linear differential equation has a unique initial value solution and any number of boundary value solutions.
In an eigenvalue problem only for certain values of the free parameter the system has boundary value solutions.
For distinction between the initial value and boundary value solutions of the same equation $ \Psi $ is used for the initial value solution and $ \psi $ for the boundary value solutions.

The equation
\be
(- \partial_t^2 + W)\Psi = \lambda \Psi
\ee
has a single solution $ \Psi_\lambda $ that satisfies the initial conditions
\be
\Psi_\lambda(t=0)=0 , \; \partial_t \Psi_\lambda(t=0)=1.
\ee
On the other hand, the operator $ - \partial_t^2 + W $ that acts on the space of functions that obey the boundary conditions $ \psi(t=0)=0 , \; \psi(t=T)=0 $ has an eigenvalue $ \lambda_n $ if and only if
\be
\psi_{\lambda_n}(0)=0 , \; \psi_{\lambda_n}(T)=0.
\ee
The determinant of the operator $ - \partial_t^2 + W $ is defined to be
\be
\det(- \partial_t^2 + W) = \prod_n \lambda_n.
\ee
It is proved in \cite{coleman} that for two different functions $  W^{(1,2)} $ with initial value solutions $ \Psi_\lambda^{(1,2)} $ the equation
\be
\det \left[ \frac{- \partial_t^2 + W^{(1)} - \lambda}{- \partial_t^2 + W^{(2)} - \lambda}  \right] = \frac{\Psi_\lambda^{(1)}(T)}{\Psi_\lambda^{(2)}(T)}
\ee
is satisfied.
As a result the normalization factor $ N $ defined by
\be
\label{colemandet}
\frac{\det(- \partial_t^2 + W)}{\Psi_{\lambda=0}(T)}=N^2
\ee
is constant.
Applying this theorem to $ n $th order difference equations \cite{benderorzag}, the equation
\be
M_{k \times k} {\vec P_k} = \lambda {\vec P_k}
\ee
has a single solution $ {\vec P}_k (\lambda) $ that satisfies the initial conditions $ P_m=0 , \; P_{m+1}=1 $.
On the other hand, the matrix $ M_{k \times k} $ that acts on $ {\vec C_k} $, which obeys the boundary conditions $ C_m=0 , \; C_{m'}=0 $ for $ m-m' \ne 0,\pm 1 $, has an eigenvalue $ \lambda_n $ if and only if $ C_m(\lambda_n)=0 , \; C_{m'}(\lambda_n)=0 $.
The determinant of the matrix $ M_{k \times k} $ is defined to be
\be
\det M_{k \times k} = \prod_n \lambda_n.
\ee
For two different matrices $ M_{k \times k}^{(1,2)} $ with initial value solutions $ {\vec P_k} $ the equation
\be
\det \left[ \frac{M_{k \times k}^{(1)}-\lambda I_k}{M_{k \times k}^{(2)}-\lambda I_k} \right] = \frac{P_{m'}^{(1)}(\lambda)} {P_{m'}^{(2)}(\lambda)}
\ee
with $ I_k $ the $ k \times k $ unit matrix, is satisfied.
This theorem can be proved following the proof given in \cite{coleman} for the differential operators.
As a result the normalization factor $ N $, defined by
\be
\frac{\det M_{k \times k}} {P_{m'}(\lambda=0)} = N^2
\ee
is constant, and the zeros of $ {P_{m'}(\lambda=0)} $ are the eigenvalues of $ \det M_{k \times k} $. 

%%%%%%%%%%%%%%%%%%%%%%%%%%%%%%%%%%%%%%%%%%%%%%%%%%%%%%%%%%%%
\chapter{\wkb Approximation}
%%%%%%%%%%%%%%%%%%%%%%%%%%%%%%%%%%%%%%%%%%%%%%%%%%%%%%%%%%%%

\label{wkbapp}
The \wkb approximation is a semi-classical approximation of the wave functions and energy levels of the \schr equation.
In this work the well knows \wkb approximation \cite{gasiorowicz,benderorzag,merzbacher} is expanded to include initial value solutions of the \schr equation.
In the \wkb approximation the solution of the \schr equation
\be
\label{wkbschr}
- \half \frac{d^2}{dx^2} \psi =\frac{1}{\hbar ^2} (E-V(x)) \psi
\ee
for small $ \hbar $ is assumed to have the general form
\be
\psi^{\wkb} = \exp \left[ \frac{1}{\hbar} \sum_{n=0}^{\infty} \hbar ^n S_n \right].
\ee
Since the \schr equation is a second order equation the general solution has two free parameters.
\bea
\psi^{\wkb} & = & A(E) e^{ i \left( \hbar ^{-1} S_0 - \hbar S_2 + \hbar ^3 S_4 - \cdot \cdot \cdot \right)}e^{ \left( S_1 - \hbar ^2 S_3 + \hbar ^4 S_5 - \cdot \cdot \cdot \right)} \non \\
& + & B(E) e^{ - i \left( \hbar ^{-1} S_0 - \hbar S_2 + \hbar ^3 S_4 - \cdot \cdot \cdot \right)} e^{ \left( S_1 - \hbar ^2 S_3 + \hbar ^4 S_5 - \cdot \cdot \cdot \right) }.
\eea
The terms $ S_n $ are found by solving equation (\ref{wkbschr}) order by order in $ \hbar $.
The explicit solution to first order in $ \hbar $ is
\bea
\psi^{\wkb} & = & \frac{A(E)}{\sqrt[4]{E-V(x)}} \; \exp \left[ i \sqrt{2} \int_0^x \sqrt {E-V(y)} \; dy\right] \non \\
& + & \frac{B(E)}{\sqrt[4]{E-V(x)}} \; \exp \left[ -i \sqrt{2} \int_0^x \sqrt {E-V(y)} \; dy \right].
\eea
For any order of the approximation two more conditions are required to determine $ A(E) $ and $ B(E) $.
The quantization of the wave functions and energy levels is a result of the boundary value conditions, which are given by the normalization condition
\be
\psi (x \ra \pm \infty) \ra 0.
\ee
The exact and approximate initial value solutions of the \schr equations have initial conditions of the form
\be
\Psi(x=0) = a , \; \frac{d}{dx}\Psi(x=0) = b.
\ee
If the potential is even then the generating function will have a well defined parity.
Therefore the initial conditions for even potentials will be either
\be
\Psi(x=0) = 1 , \; \frac{d}{dx}\Psi(x=0) = 0
\ee
or
\be
\Psi(x=0) = 0 , \; \frac{d}{dx}\Psi(x=0) = 1
\ee
depending on the parity of the states.
The result of this approach is the initial value \wkb approximation.

%%%%%%%%%%%%%%%%%%%%%%%%%%%%%%%%%%%%%%%%%%%%%%%%%%%%%%%%%%%%
\newpage
\addcontentsline{toc}{chapter}{Bibliography}
\bibliography{qes}
%%%%%%%%%%%%%%%%%%%%%%%%%%%%%%%%%%%%%%%%%%%%%%%%%%%%%%%%%%%%

%%%%%%%%%%%%%%%%%%%%%%%%%%%%%%%%%%%%%%%%%%%%%%%%%%%%%%%%%%%%
\end{document}